\documentclass[fleqn,usenatbib]{mnras}

\usepackage[T1]{fontenc}

\DeclareRobustCommand{\VAN}[3]{#2}
\let\VANthebibliography\thebibliography
\def\thebibliography{\DeclareRobustCommand{\VAN}[3]{##3}\VANthebibliography}

\usepackage{graphicx}	% Including figure files
\usepackage{amsmath}	% Advanced maths commands
\usepackage{amssymb}
\usepackage{gensymb}

\usepackage{subcaption}
\usepackage{geometry}
\usepackage{float}
\usepackage[font=small,labelfont=bf]{caption}
\usepackage{multicol}
\usepackage{xcolor}

\usepackage{newtxtext,newtxmath}

\title[Hycean GCMs]{General Circulation Models of Hycean Worlds}

\author[E. F. L. Barrier et al.]{
Edouard F. L. Barrier,$^{1}$\thanks{E-mail: efxlb2@ast.cam.ac.uk}
Nikku Madhusudhan$^{1}$\thanks{E-mail: nmadhu@ast.cam.ac.uk}
\\
$^{1}$Institute of Astronomy, University of Cambridge, Madingley Road, Cambridge, CB3 0HA, UK
}

\date{Accepted 2025 October 30. Received 2025 October 30; in original form 2025 July 04}

\pubyear{2025}

\begin{document}
\label{firstpage}
\pagerange{\pageref{firstpage}--\pageref{lastpage}}
\maketitle

\begin{abstract}

Sub-Neptunes represent the current frontier of exoplanet atmospheric characterisation. A proposed subset, Hycean planets, would have liquid water oceans and be potentially habitable, but there are many unanswered questions about their atmospheric dynamics and 3D climate states. To explore such climates in detail, we report a General Circulation Model (GCM) for Hycean worlds, building on a modified version of the ExoCAM GCM. Considering the temperate sub-Neptune K2-18~b as a Hycean candidate, we implement GCMs with different surface pressures and albedos. We find dynamical structures similar to those of tidally-locked terrestrial planets as `slow rotators' with either one equatorial or twin mid-latitude zonal jets. We see moist convective inhibition that matches high resolution models, although in hotter cases the inhibited zone is subsaturated. When imposing a top-of-the-atmosphere (TOA) Bond albedo ($A_b$) by modifying the incident stellar flux, we find that the threshold for K2-18~b to not enter a runaway greenhouse state is $A_b~\geq~0.55$ for a 1 bar atmosphere, consistent with previous studies, and $A_b~\geq~0.8$ for a 5 bar atmosphere. However, a more realistic treatment of the albedo, by modelling scattering within the atmosphere using an enhanced Rayleigh parametrisation, leads to lower lapse rates and stronger thermal inversions. We find that 1 bar atmospheres are stable for an albedo of $A_b~\geq~0.27$, 5 bar atmospheres for $A_b~\geq~0.35$, and 10 bar atmospheres for $A_b~\geq~0.48$. Moderate albedos such as these are typical of the solar system planets and the required scattering is consistent with observational constraints for K2-18~b, supporting its plausibility as a Hycean world.

\end{abstract}


%%%%%%%%%%%%%%%%%%%%%%%%%%%%%%%%%%%%%%%%%%%%%%%%%%

%%%%%%%%%%%%%%%%% BODY OF PAPER %%%%%%%%%%%%%%%%%%

\section{Introduction}

The search for life beyond Earth is one of the most compelling aspects of the human condition. In the field of exoplanets, this means searching for habitable planets and biosignatures \citep{Cockell2016, Meadows2018}. Traditionally this has focused on Earth-like planets, under the reasonable motivation that Earth is the one place we have found life so far. However, characterisation of Earth-like planets would require a large investment of observation time with current facilities, if it is possible at all \citep{Fauchez2022,Meadows2023}. This is a good motivation to consider whether other classes of planets also might possess the capacity to be habitable. One such proposed class is that of Hycean planets \citep{Madhusudhan2021}. Hycean planets have H$_2$-rich atmospheres and a high H$_2$O mass fraction, leading to very deep liquid water oceans \citep{Rigby2024a}. Inhabiting a region of exoplanet mass-radius space where internal structure solutions are often degenerate, they can be significantly larger than Earth - canonically, up to 2.6 Earth radii and 10 Earth masses. The combination of these larger sizes, and the increased scale heights of their hydrogen-rich atmospheres, mean that they would be excellent prospects for atmospheric characterisation using current facilities. It was predicted that biosignatures would be detectable at ppm quantities with the JWST \citep{Madhusudhan2021}.

K2-18b \citep{Cloutier2019} is the archetypal Hycean candidate, a designation motivated initially by internal structure models which allowed a liquid water ocean \citep{Madhusudhan2020}. JWST Cycle 1 observations found no signs of water vapour suggested by HST observations \citep{Benneke2019,Tsiaras2019}, but they did reveal of the presence of CH$_4$ (to 5$\sigma$) and CO$_2$ (to 3$\sigma$) at roughly 1\% levels \citep{Madhusudhan2023b}. This, coupled with the non-detection of NH$_3$ and CO, has been interpreted as suggesting the presence of a surface at relatively low pressures \citep{Madhusudhan2023a}. Given the Mass-Radius constraints on the internal structure, such a surface would be consistent with liquid water and K2-18b could be a Hycean planet. 

However, there are other possible internal structure solutions \citep{Madhusudhan2020}, and a number of these alternatives have been put forward to explain the observed atmospheric composition. A reanalysis of photochemical networks in \citet{Wogan2024} suggested that a Mini-Neptune scenario is as plausible as a Hycean, however \citet{Cooke2024} indicated that the Mini-Neptune scenario is less consistent with the retrieved abundances compared to a Hycean, although no case (bar an inhabited Hycean with an arbitrary CH$_4$ flux) can match the observed constraints on all five molecules. \citet{Shorttle2024} proposed that K2-18b might have a magma ocean, although it has been argued that this is also inconsistent with the observed atmospheric composition \citep{Glein2024,Rigby2024b}. \citet{Luu2024} also investigated the chemical signatures that a supercritical water layer might cause and suggested that the CH$_4$/CO$_2$ and CO/CO$_2$ ratios should act as constraints on the existence of such a layer. 

Recently, JWST MIRI observations found potential evidence for DMS and/or DMDS at 3$\sigma$ \citep{Madhusudhan2025}, building on a tentative indication of DMS using near-infrared data \citep{Madhusudhan2023b}. This is remarkable as there is no known abiotic pathway that could produce the observed amounts of DMS, raising the prospect of an inhabited - and Hycean - K2-18b. A follow-up analysis from \citet{Welbanks2025} proposed that six other molecules could provide comparable fits to the data, while a more extended search \citep{PC2025} found that one or more of three molecules, including DMS, provide the best explanation to all the data available.  Overall, the question of the nature of K2-18 b is not yet settled - future observations should help provide more constraints - but the Hycean option remains a distinct possibility.

In light of this, it becomes imperative to know as much as possible about Hycean planets. One important set of unanswered questions revolves around their climate state as whole. So far, most atmosphere modelling has been done in 1D \citep[e.g.][]{Madhusudhan2021, Innes2023, Jordan2025}, with some high-resolution work investigating small domains in 3D (e.g. \citet{Leconte2024}). There has not been any work which studies Hycean planets as the 3D objects they are, which severely limits the detail in which these objects - particularly their global, inherently three-dimensional processes - can be studied. Key climate features which are therefore uncertain include the geographical variations in temperature and moisture, the dynamical structure and wind patterns, and cloud and precipitation patterns. Crucially for the case of K2-18b, it is also not clear where and when the runaway greenhouse threshold develops on Hycean planets when they are simulated in 3D. 

This is important because the strongest argument against K2-18b being a Hycean planet has not been an atmospheric chemistry argument but rather a climate one. Due to the strong greenhouse effect of H$_2$ \citep{PierrehumbertGaidos2011} and the presence of the liquid water surface reservoir, a Hycean K2-18b might enter into a runaway greenhouse state which would destroy its habitability in the absence of an adequate albedo \citep{Innes2023, Leconte2024}. The phenomenon of moist convective inhibition amplifies this effect. For a high mean molecular weight condensible in a lighter background atmosphere, moist convection will be inhibited if the mass mixing ratio of the condensible crosses the "Guillot Threshold" \citep{Guillot1995}. This leads to substantially higher surface temperatures and reduces the maximum instellation for which a stable climate is possible. Understanding the precise parameter space in which a Hycean K2-18b could exist is thus of paramount importance for assessing its possible nature, and so the prospects for habitability. 

To answer these questions about the climate and atmospheric structure of a Hycean K2-18b, we use a 3D General Circulation Model (GCM). GCMs are models which simulate the dynamics and various physical processes happening across the whole atmosphere, allowing us to model these atmospheres fully self-consistently and with a minimum of approximations \citep{Pierrehumbert2010,Showman2013}. They consist of, at a minimum, a dynamical core solving some form of the fluid equations of motion as well as some physics parametrisations. GCMs are widely used to model many aspects of exoplanet atmospheres such as their dynamical structure \citep{Innes2022}, photochemistry \citep{Cooke2023},  clouds \citep{Charnay2021}, and higher-level considerations of habitability such as demarcating the Inner Habitable Zone \citep{Kopparapu2016}. 
In our case, we use the ExoCAM GCM \citep{Wolf2022}, an offshoot of NCAR's Community Systems Earth Model \citep{Neale2010}. We have added an updated convective parametrisation, presented more fully in \citet{Barrier2025}, which allows us to simulate moist convective inhibition and the runaway greenhouse threshold. This is the first GCM simulation of a Hycean atmosphere and allows us to investigate the global climate, the dynamical and temperature structure, and particularly the behaviour of clouds and convection in the substellar region. 

There have been a number of recent works which have investigated the fundamental behaviour of moist convective inhibition and how this affects the habitable zone of Hycean planets. The linear analysis of \citet{Leconte2017} confirmed moist convection inhibition as a process and showed that layers stable to moist convection were also stable to double-diffusive instabilities and so should remain superadiabatic. \citet{Innes2023} applied a 1D cloud-free radiative-convective model (heat can be transported by radiation and convection) to find values for the runaway greenhouse threshold. For example, they found that the threshold for a 1 bar $H_2$ atmosphere orbiting an M-dwarf is a planet-averaged 115 Wm$^{-2}$, as compared to an estimated instellation of 342 Wm$^{-2}$ for K2-18b in the absence of clouds or hazes \citep{Cloutier2019}.

\citet{Leconte2024} studied moist convective inhibition using a 3D high-resolution Convection Permitting Model (CPM), as have \citet{Habib2024b} and \citet{Seeley2025}. These CPMs have resolutions of $O(100m-1km)$ and can resolve convection explicitly without needing sub-grid parametrisations, making them appealing tools to verify whether a theorised effect such as moist convective inhibition occurs in real life. These studies reach a common conclusion: moist convective inhibition exists as a phenomenon, either in the free troposphere as in the mini-Neptune configurations in \citet{Leconte2024} and \citet{Habib2024b} or in the surface-adjacent region like in \citet{Seeley2025}. Heat and transport appears to be transported through the stable layer by small-scale turbulence, a departure from the work of \citet{Innes2023}. As a result, \citet{Leconte2024} find slightly more forgiving constraints on the runaway greenhouse threshold. For a 1 bar atmosphere on K2-18b, the threshold is a planet-average instellation of between 171 and 137 $Wm^{-2}$, corresponding to a Bond albedo of 0.5 to 0.6.

Where these works disagree is on the amount of this turbulence and consequently on the lapse rate through the stable layer, with important impacts on the runaway greenhouse transition and overall climate state. Notably, \citet{Habib2024b} find much weaker vertical transport than \citet{Leconte2024}, which might correspond to a higher lapse rate in the inhibition region. It is hard to compare to \citet{Seeley2025} as they do not directly investigate this, but they do find very high lapse rates of up to 120 K/km for portions of the surface-adjacent inhibition layer.

For all the strengths of these models, they do have some limitations. CPMs are too computationally expensive to run on the whole planet and so are run on a restricted spatial scale, for example 128km by 128km in \citet{Leconte2024}. Normally, perfect heat distribution across the planet is assumed such that the top of the CPM domain receives the planet-averaged incident flux, and they can be conceptualised as a box placed at some point on the dayside of the planet. This lack of global coverage means that they cannot capture the atmospheric circulation, the geographic variations in temperature and the exact behaviour of other processes, and cannot give a fully self-consistent climate. Particularly for processes such as convection, which is mostly driven by stellar shortwave heating of near-surface air, assuming globally uniform behaviour can be a limiting assumption. This motivates a GCM study, so that we can model the atmosphere of a Hycean K2-18b simulation fully self-consistently. It is useful to use insights from the 3D models above in convective parametrisations, as we have done in \citet{Barrier2025}.

Beyond investigating the climate states and runaway greenhouse threshold of a Hycean K2-18b, we would also like to investigate and understand the dynamical structure. There exists a substantial body of work outlining what dynamical structure we might expect to find. Some key characteristics of K2-18b are particularly relevant for this. It has a larger radius than Earth, which matters for quantities such as the dimensionless Rossby radius. It has an orbital period of 32.94 days \citep{Benneke2019} and is presumably tidally locked, and receives a very similar instellation to the Earth of 1368 Wm$^{-2}$. For our canonical cases, we assume atmospheres of 1 and 5 bar surface pressure, meaning that surface temperature variations are still possible. 

The slow rotation speed means that the Coriolis force is relatively weak when it comes to influencing the large scale dynamics. We might expect to be in a `tropical' regime characterised by small horizontal heat gradients \citep{Vallis2017}. The tidally-locked nature of the planet means we also expect at least an amount of overturning circulation, with large scale upwards motions at the substellar point and a corresponding divergence away from it. It is not completely clear what dynamical wind structure we expect. Many simulations of slow rotating exoplanets show super-rotating equatorial jets \citep{Showman2013, Hammond2020}, but some also show mid-latitude jets that can be sustained either by meridional temperature gradients or by turbulent eddies leading to momentum convergence \citep{Pierrehumbert2019, Innes2022}. In hotter cases with significant fraction of a condensible component, we might expect a more barotropic circulation with a deep but weak global Hadley cell \citep{Pierrehumbert2016}.

In this work we set out to explore the possible climate states of a Hycean K2-18b. We use the ExoCAM GCM with the convective parametrisation from \citet{Barrier2025} and a small number of additional modifications. We seek to impose 3D-informed constraints on the albedo required for a stable Hycean climate with a range of surface pressures. We explore two different prescriptions for inducing this albedo: a simple reduction in the top of atmosphere (TOA) incident flux, and parametrising hazes by increasing the amount of H$_2$ Rayleigh scattering. For both of these prescriptions, we run suites of GCM simulations where we vary the surface pressure (taking values of 1 and 5 bar) and also the induced albedo. In Section \ref{sec:methods} we describe the details of the GCM and the parametrisations we use. We start by studying one particular GCM case in detail in Section \ref{sec:casestudy}. Taking a 1 bar Hycean case with an imposed TOA A$_b$ of 0.6, we focus on the general temperature and dynamical structure and take a close look at the behaviour of convection near the substellar point, allowing us to see how our global 3D model matches up against high-resolution CPM results. In Section \ref{sec:r2_ab} we present the rest of the imposed TOA cases, and describe the effects of changing the surface pressure and induced albedo. In Section \ref{sec:r3_rs} we implement a more realistic source of albedo by parametrising hazes as increased Rayleigh scattering. We present the results, and again see what climate states and runaway greenhouse thresholds we find. We summarise all our results and discuss their implications in Section \ref{sec:summary}.

\section{Methods}

In this section we describe the numerical models used in this work. We begin by providing an overview of the General Circulation Model (GCM). We then describe in some more detail the convection scheme, considerations to simulate a Hycean world, and our different methods to account for an albedo.

\label{sec:methods}

\subsection{The ExoCAM GCM}
We use the ExoCAM GCM \citep{Wolf2022} as our starting point for these simulations. ExoCAM is a specially developed GCM for use in terrestrial exoplanet settings, and is an independently curated branch of NCAR's Community Earth Systems Model (CESM), version 1.2.1 \citet{Neale2010}. It has been used for a wide variety of applications including GCM simulations of terrestrial exoplanets and mini-Neptunes \citep{Sergeev2022, Barrier2025}.

The ExoCAM dynamical core uses a finite-volume scheme \citep{Lin1996} to solve the primitive equations of meteorology \citep{Vallis2017} on an Arakawa C grid. Due to the relatively cool temperatures and consequently moderate temperature contrasts in our simulated atmospheres, we expect that the primitive equations remain an appropriate approximation as discussed in \citet{Mayne2019, Charnay2021, Innes2022}.
We impose both a 4th-order velocity damping and a 2nd-order velocity diffusion for the top layers, leading to an implicit numerical sponge layer.

We run our simulations with a horizontal resolution of 72 by 45 cells, equivalent to cells being $5\degree$ longitude by $4\degree$ latitude. We use 61 vertical levels spanning 4 orders of magnitude so that cases with a 1 bar surface have a model top at 10 Pa and those with a 5 bar surface have a 50 Pa model top. These vertical layers are mostly equally spaced in log pressure, but are concentrated towards the near-surface layers: 11 of the layers are between the surface pressure $P_s$ and $0.9P_s$. The few layers near the surface have $\Delta p/p \approx 0.003$, with the pressure thicknesses increasing in the free troposphere. This gives us the required vertical resolution to resolve near surface effects, notably the presence of a stable non-convecting layer \citep{Innes2023,Seeley2025}. 

In this work we conduct GCM simulations for surface pressures between 1 and 10 bar. This is consistent with  observational constraints on K2-18b, as the non-detection of NH$_3$ in the atmosphere is best explained by a thin atmosphere, e.g. total pressure below a few tens of bars  \citep{Madhusudhan2023b, Cooke2024}. Recent works on Hycean climates have also considered pressures between 0.1 and 10 bar for such models, motivating our choice of pressure range in the present work \citep{Innes2023,Leconte2024,Kuzucan2025}. We note that most of the simulations we run are for 1 bar and 5 bar surface pressures for computational convenience. Atmosphere spin up times rapidly grow for thicker atmospheres, because of the larger dry mass of the atmosphere, and the reduced energy influx reaching the deep atmosphere. Additionally, given our search for stable Hycean climates, thicker atmospheres have cooler upper layers and so longer radiative timescales. We have thus prioritised the ability to first test two different methods of inducing an albedo, and also to examine multiple induced albedos for each of these methods. We do also simulate a 10 bar atmosphere in Section \ref{sec:r3_rs} to explore constraints for thicker atmospheres.

A number of preceding exoplanet GCM simulations using the CESM finite volume dynamical core have reported issues with angular momentum (AM) conservation, especially for slowly rotating planets \citep{Lebonnois2012, Lauritzen2014}. Given that we assume that K2-18b is tidally locked and so has a rotation period of 32.94 days, our simulations are very much in this regime. This effect is amplified by the fact that we have a flat surface with no orography, reducing the amount of surface torque on the atmosphere. To handle this, we follow \citet{Barrier2025} in implementing the recommended changes from \citet{Toniazzo2020}. They found the main source of AM non-conservation was in the solver for the shallow water equations, and developed a correction for this as well as a fixer to conserve global angular momentum in the dynamical core. 

We use the ExoRT \textit{n68equiv} radiative transfer scheme as detailed in \citet{Wolf2022}. It is a two-stream correlated-k distribution scheme with 68 spectral intervals from 0.238-infinity $\mu m$, and uses the equivalent-extinction method to handle overlapping gases \citep{Amundsen2017}, which easily handles many overlapping species in each spectral interval. It includes gas absorption for $H_2O$, $CO_2$, and $CH_4$ from the HITRAN 2016 database \citep{Gordon2017}, self and foreign water vapour continuum using MT-CKD version 3.2. In addition to $N_2-N_2$ and $H_2-H_2$ CIA, $CO_2-CO_2$ \citep{Wordsworth2010} and $CO_2-H_2$ and $CO_2-CH_4$ \citep{Turbet2020} CIA pairs are also included. Liquid and ice cloud droplets are treated as Mie scattering particles and cloud overlap is treated using the Monte Carlo Independent Column Approximation with maximum cloud overlap. Rayleigh scattering is included using the parametrisation of \citet{Vardavas1984}, including the effects of $N_2$, $CO_2$, $H_2O$, and $H_2$.

The CAM 4 moist physics package \citep{Neale2010} predicts the water vapour, liquid cloud, and ice cloud concentrations using a Sunqvist-style bulk microphysical parametrisation \citep{Sundqvist1988, Zhang2003}. The cloud cover and precipitation fluxes are diagnosed from the above variables, and there is a fixed cloud droplet number concentration of $1.5\times10^{8} m^{-3}$. There are three types of cloud diagnosed: marine stratus clouds, convective clouds (which we only allow to exist if water has condensed in the convective scheme), and layered cloud if the grid-box relative humidity (RH) is high enough.

Convection is by default handled in ExoCAM by the \citet{Zhang1995} deep and shallow \citep{Hack1993} moist convection schemes. The Zhang-Macfarlane deep convection scheme is a mass-flux scheme and uses a plume ensemble approach, with updrafts and downdrafts triggered by conditional instability diagnosed from entraining plume ascent. The mass fluxes are determined by destroying two-thirds of the Convective Available Potential Energy (CAPE) every two hours. The Hack shallow convection handles cumulus convection occurring in 3 layers or less and in some respects works similarly to a number of convective adjustment schemes, iterating up through the atmosphere until convective instability is eliminated.

\subsection{Changes to convection scheme and turbulent physics}

The deep \citep{Zhang1995} and shallow \citep{Hack1993} moist convection schemes have been reworked from the default CAM4 parametrisations. This is presented in detail in \citet{Barrier2025} and allows the convection scheme to handle a series of non-Earthlike convective modes, notably convection in a low molecular weight atmosphere, as well as convection in a non-dilute atmosphere. We review some key details here.

The first major change is the criterion of convective initialisation. We use virtual potential temperature \citep[e.g.][]{Habib2024a} to diagnose instability to dry convection, and a moist convection criterion which accounts for the buoyancy effect of water vapour from \citet{Leconte2017}. We allow more flexibility in the choice of convection start layer and also allow for the existence of two separate deep convection zones. We use entropy as a governing variable in the plume ascent and when calculating entrainment rates. We make a further series of changes to the plume ascent calculations to ensure accuracy across all our different convective modes. We make a few changes to the Hack shallow convection scheme, changing its initiation criterion and allowing it to handle dry convection.

We also makes changes to the planetary boundary layer (PBL) turbulent mixing processes. Virtual potential temperature $\theta_v$ as defined in Equation \ref{eq:theta_v} is used instead of potential temperature $\theta$ when calculating layer stabilities, allowing for compositional stratification. 

\begin{equation}
    \Theta_v = T (1 - \bar{\omega} q )e^{- \int_{p_0}^{p} R_{m} / C_{p,m} \,d \ln p} \label{eq:theta_v}
\end{equation}

$\bar{\omega}$ is the reduced mean molar mass difference $\bar{\omega} = \frac{\mu_v - \mu_d}{\mu_v}$. $C_{p,m}$ is the composition adjusted heat capacity and $R_{m}$ is the composition adjusted gas constant. The $_{m}$ subscripts mean that these values are constructed from a mass-weighted average of the individual gas species values, $C_{p,m} = (1-q) C_p^d + q C_p^v$ and $R_{m} = R_g ( \frac{1-q}{\mu_d} + \frac{q}{\mu_v} )$, where the $_d$ subscript refers to the combination of dry (non-condensing) gases and $_v$ refers to a potentially varying condensible component ($H_2O$ is this case). The advantage of this change is that PBL heat and moisture fluxes increase when the PBL is unstable, as originally diagnosed with $\theta$ decreasing with height. Using $\theta_v$ instead allows for cases where there is a significant destabilising temperature gradient, but also a stabilising moisture gradient.

We adjust the value of the minimum vertical mixing coefficient $K_{zz,min}$. This is set to a default minimum $0.01 m^2 s^{-1}$ and usually relatively unimportant, but because of our compositional stratification (varying H$_2$O in a H$_2$ background) it becomes an important variable and usually sets the $K_{zz}$ in the near-surface dayside regions. We set it to $0.08 m^2 s^{-1}$ following the minimum $K_{zz}$ found in \citet{Leconte2024}, although we note that it is an important free parameter whose variability is not currently well understood.

\subsection{Parametrising albedos and hazes}

One of the main aims of the paper is to see for which levels of imposed albedos or hazes K2-18b might be able to sustain a liquid water surface ocean. We use two different prescriptions for setting the albedo, which in turn reduces the surface temperature.

The first method is to simply reduce the instellation reduced by the planet by a given fraction. We achieve this by imposing an albedo $A_b$ such that the incoming stellar flux $F_{inc} = (1-A_b)F_{irr}$, with $F_{irr}=1368 Wm^{-2}$ the expected instellation that K2-18b receives given its orbital separation and the properties of its parent star K2-18. We refer to this as the top of the atmosphere (TOA) albedo. We note that the true albedo of the planet (how much incoming starlight is reflected) will be slightly higher than this as $A_i$ does not account for reflections, for example from clouds. However this difference is small as the relatively high optical depth of even a 1 bar atmosphere means that the majority of the incoming flux is absorbed.

The second method is a commonly used parametrisation \citep{Madhusudhan2021, Piette2020} and involves changing the amount of Rayleigh scattering to mimic the effects of hazes, as most show similar behaviour with increasing opacities towards the bluewards end of the spectrum \citep{Pinhas2017}. This mimicks a haze with only scattering and no absorption, like the extra-reflective soot hazes in e.g. \citet{Malsky2025}). We implement the parametrisation by increasing the optical depth attributed to Rayleigh scattering by given enhancement factors. With $1\times$ scattering representing no added hazes, we work with enhancement factors $E$ of $100\times$, $1000\times$, and $10000\times$. For example, taking a layer with an $H_2$ column density of $n_{H_2}$ molec m$^{-2}$, the haze optical depth of the layer $d\tau_h(\lambda) = \sigma_{ray,H_2}(\lambda) n_{H_2} E$. 

Continuing our approach of a simple haze parametrisation, we assume that the haze is distributed uniformally geographically. This is an approximation, as photochemical hazes would form on the dayside before spreading to the nightside and the terminators, and so geographically uniform hazes will underestimate their net cooling effect. We do not, however, assume a constant vertical value of the enhancement factor $E$. Doing this leads to very high optical depths at high pressures, as $n_{H_2}$ increases linearly with pressure. This is in contrast to more realistic haze models  \citep[e.g.][]{Steinrueck2025,Malsky2025}, where this increase in $d\tau/d\ln{P}$ with higher pressure is not seen. We decrease the enhancement factor $E$ inversely for pressures greater than $p_0=100$ mbar, i.e. $E=E_0 (p_0/p)$ for $p>100$ mbar. This is a more realistic implementation of hazes and allows us to avoid numerical instabilities resulting from the high scattering at high pressures. The results presented later in the paper should be understood with this kept in mind: a 1000x haze enhancement factor will have a somewhat lesser effect on the climate than found in other work. Given the number of possible free parameters in parametrising the haze, it is possible that differences in haze parametrisation could lead to substantially different climate states. We believe that a simple parametrisation is best justified for an initial investigation into the effects of hazes.

\section{Results I - Reference case}
\label{sec:casestudy}

\begin{table}
\centering
\begin{tabular}{cc}%{|c|c}

\hline
\textbf{Planet Properties} & \\
Radius / $m$ & $1.66 \times 10^7$ \\
Rotation period / days & 32.94 \\
Gravity / $m s^{-2}$ & 12.4 \\
Instellation / $W m^{-2}$ & 1368 \\
Internal Temperature $T_{int}$ / $K$ & 30\\
Ocean albedo & 0.00 \\
\hline
\textbf{Atmospheric composition} & \\
H$_2$ dry VMR  & 0.83167 \\
He dry VMR  & 0.142 \\
\textbf{CH$_4$} dry VMR  & 0.0182 \\
\textbf{CO$_2$} dry VMR  & 0.00813 \\
\hline
\end{tabular}
\caption[Simulation parameters held constant across our Hycean K2-18b runs]{Simulation parameters held constant across our Hycean K2-18b runs. Planetary parameters taken from \citet{Benneke2019} and \citet{Cloutier2019}. $T_{int}$ is set at 30K based on considerations from \citet{Valencia2013}. Atmospheric composition from JWST observations in \citet{Madhusudhan2023b}, assuming a solar $H_2$/He split. A surface albedo of 0 is set to simplify results. 'Dry VMR' refers to the volume mixing ratio of the gas relative to a dry mixture, i.e. all the atmospheric constituents except for H$_2$O.}\label{tbl:k218b_params}
\end{table}

\begin{table*}
\centering
\begin{tabular}{cccccc}%{|c|c|c|c|c|c}

\hline \hline
\textbf{Run Name} & \textbf{$P_s$ / bar} & \textbf{$S$ / $S_0$} & \textbf{Imposed $A_b$} & \textbf{Run time / days} & \textbf{Stable Climate?} \\
\hline
1b-548w & 1 & 0.403 & 0.6 & 35,000 & \color{green} \checkmark \\ %548
1b-582w & 1 & 0.428 & 0.575 & 35,000 & \color{green} \checkmark \\ %582.25
1b-616w & 1 & 0.453 & 0.55 & 35,000 & \color{green} \checkmark \\ %616.5
1b-650w & 1 & 0.478 & 0.525 & 23,000 & \color{red} $\times$ \\ %650.75
1b-685w & 1 & 0.503 & 0.503 & 16,000 & \color{red} $\times$ \\ %685
\hline
5b-274w & 5 & 0.201 & 0.8 & 55,000 & \color{green} \checkmark \\ %274
% 5b-342w & 5 & 342.5 & 0.75 & 55,000 & \color{green} \checkmark \\
5b-411w & 5 & 0.302 & 0.7 & 30,000 & \color{red} $\times$ \\ %411
\hline \hline
\end{tabular}
\caption[Simulation details of $A_b$ suite]{Simulation details of the first suite of runs with modified instellation. A run which has converged is one which has reached a stable climate state, the ones which have not are in a runaway greenhouse state. Run times are listed to the closest 1000 days.}\label{tbl:ab_runs}
\end{table*}

In this work, we explore the possible dynamical regimes of a Hycean world by running successive series of GCM simulations of K2-18b, varying the surface pressure and different prescriptions for inducing an albedo.

In our first set of runs we simply modify the incident stellar flux by the factor $1-A_b$. We refer to this as the top-of-atmosphere (TOA) albedo. The runs carried out are listed in Table \ref{tbl:ab_runs}. We consider that the climate states have converged when the kinetic energy of the atmosphere reaches a steady value, the net emergent longwave flux reaches a steady value such that the OLR is equal to the incident stellar irradiation, and the globally-averaged surface temperature is steady. In cases where the state does not converge, the climate is in a runaway greenhouse state: it is generally obvious when this happens as the surface temperature ends up well above boiling point. Results are presented averaged over the last 1000 days of simulation time.

Before focusing on the runaway greenhouse threshold of a habitable K2-18b and the whole suite of induced TOA albedo cases, we first choose to take a deeper look at the climate and dynamical structure of a case that is securely in steady state and far from runaway greenhouse conditions: that of a 1 bar atmosphere with a bond albedo $A_b$ of $0.6$, equating to a substellar point instellation of $548Wm^{-2}$. We describe first the temperature and radiative structure of the planet, then its dynamical structure. We finish off with a detailed look at the behaviour of convection and other processes near the surface around the substellar point, which allows us to compare our results to that of the high-resolution CPMs of \citet{Leconte2024}, \citet{Habib2024b}, and \citet{Seeley2025}. 

The key planetary parameters are kept constant throughout our runs and are listed in Table \ref{tbl:k218b_params}. This includes our (dry) atmospheric composition: we take the best fit observed abundances from the canonical one-offset case from \citet{Madhusudhan2023b}. We also assume that K2-18b is tidally locked, as is very likely if it is a Hycean: \citet{Tobie2019} found that the tidal dissipation quality factors $Q$ for ice-rich planets can reach as low as $\approx 2$ depending on the tidal frequency. This could plausibly lead to increased internal heat fluxes for closer-in, eccentric Hyceans, but K2-18 b is far enough from its host star that we do not expect this to be important here \citep{Livesey2025}.

All of our cases are initialised in a cold start configuration with an at rest, isothermal 300K atmosphere. We use ExoCAM's slab ocean configuration, which represents the ocean as a single layer with a depth of 100m, broadly corresponding to the depth of the wind-mixed layer on Earth. This model does neglect the influence of ocean dynamics and heat transport on the climate, which can act to increase heat transport away from the substellar region \citep{Hu2014,Yang2019b, DelGenio2019}. However, in hotter conditions the atmosphere dominates the heat transport and so a dynamic ocean would not be expected to affect the onset of the runaway greenhouse transition \citep{Yang2019a}.

\subsection{Temperature and Cloud Structure}

\begin{figure*}
    \centering
    \includegraphics[width=0.95\textwidth]{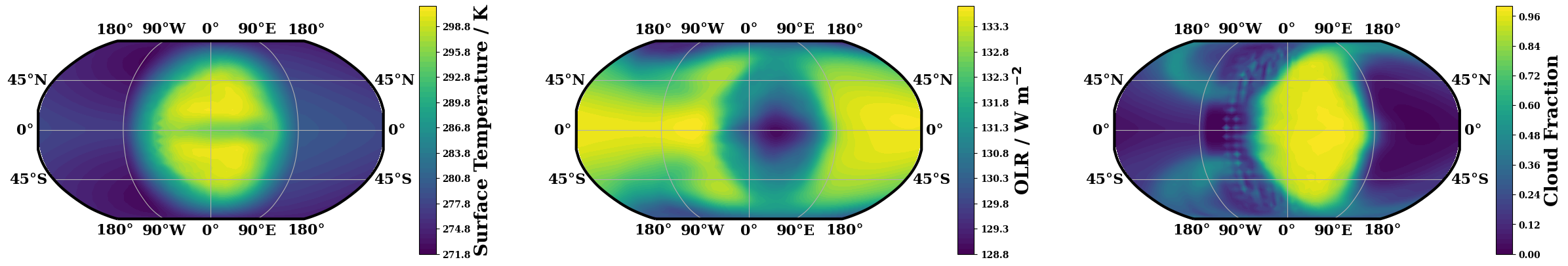}
    \caption{Various climate features in our imposed TOA $A_b=0.6$ case. Left: The surface temperature in our simulations. The higher temperatures on the dayside are immediately apparent, steady increasing going towards substellar point, with the exception of a slightly cooler equatorial band around the substellar band. The maximum temperature difference is about 30 K, and the coldest regions are approximately at the freezing point of water. Middle: The outgoing longwave radiation (OLR) representing the net thermal energy emitted by the planet. It is balanced by the incoming stellar radiation. The global OLR is generally very uniform, with the exception of a very slight decrease in a colder regions in the nightside gyres and and a small decrease around the substellar region as a result of the clouds there. However, this only leads to a difference of 5\% in OLR. Right: The global cloud cover. Clouds are concentrated on the substellar point and the region east of it, and streams of clouds entrained eastwards by the high-latitude jets can also be seen. }
    \label{fig:casestudy_ts_olr_cld}
\end{figure*}

The surface temperature map is shown in the left panel of Figure \ref{fig:casestudy_ts_olr_cld}. There is a moderate temperature difference between the dayside and nightside, with the contrast between the hottest and coldest points on the surface about 30 K. By construction, the overall temperatures are clement: the bulk of the dayside is at about 300K, and the equatorial regions of the nightside around 280K. The whole surface stays above the freezing point of water, although only barely above it in the cold gyres in the western hemisphere nightside.

\begin{figure}
    \centering
    \includegraphics[width=0.48\textwidth]{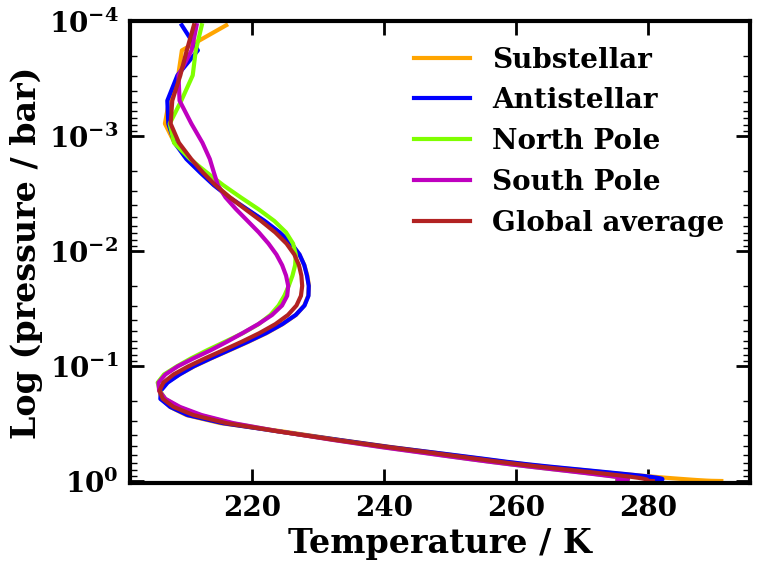}
    \caption{Pressure-Temperature profiles for the imposed TOA $A_b=0.6$ case at a range of locations across the surface: the Substellar and Antistellar points, the North and South poles, and a global average. A significant temperature inversion can be between 10 and 100 mbar driven by CH$_4$ and CO$_2$ absorption. There are  minimal temperature variations throughout the stratosphere and free troposphere, as expected from the planet largely being in a weak temperature gradient (WTG) regime. However, near the surface we see significant temperature gradients emerge.}
    \label{fig:casestudy_PTs}
\end{figure}

Despite these surface temperature variations, the rest of the atmosphere shows very few geographical variations in temperature as shown in Figure \ref{fig:casestudy_PTs}, which shows the Pressure-Temperature profiles at a number of locations on the planet, as well as a global average. There is a stratospheric temperature inversion of about 25K with the tropopause at around 0.2 bar, but the maximum horizontal temperature variations in the stratosphere and free troposphere are about 5K and increased temperature contrasts only happen in the few hundred hPa closest to the surface. This can be understood as the planet largely being in a weak temperature gradient (WTG) regime as defined in \citet{Pierrehumbert2016}. 

The WTG parameter $\Lambda$ is defined as $\Lambda = \sqrt{gD}/\Omega R_p$, with $D$ the characteristic depth scale of the wind structure. Approximating $D$ as the scale height $RT/g$, $\Lambda \approx \sqrt{RT}/\Omega R_p$. In this case, $\Lambda \approx 11$, showing that the radius of influence of steady-state response to localised heating is much greater than the radius of the planet. This is by itself not enough for the planet to necessarily be in a WTG state, though, as strong pressure gradients can be balanced by strong wind gradients \citep{Pierrehumbert2016,Vallis2017}. The two requirements to be in a WTG regime are in fact $\Lambda >> 1$ and also $\Lambda >> Ro$. In our case, the Rossby number $Ro = U/2\Omega R_p \approx 1$ so this condition is met as well, and the free tropospheric temperature gradients are minimal as expected. However, the steady-state zonal momentum equation on which the WTG derivation depends does not include frictional forces. Near the surface it cannot be expected to hold, and the nightside near-surface air cools radiatively, resulting in the surface temperature differences seen. 

The globally uniform free troposphere temperatures explain the outgoing longwave radiation (OLR), as shown in the middle panel of Figure \ref{fig:casestudy_ts_olr_cld}. There are very few horizontal variations in OLR, the greatest being a 5\% dip around the substellar region due to increased cloud cover there. This is because the total optical depth of the atmosphere is $\gtrsim$ 1 due to the high concentrations of CO$_2$ and CH$_4$ and the H$_2$-H$_2$ collisional broadening. This is in contrast to e.g. a 1 bar atmosphere with an Earth-like composition, and means that the photosphere is located around the upper troposphere and so variations in surface temperature have only a small impact on the OLR. This increased optical depth goes some way to explaining the reduced surface temperature differences compared to other simulated of tidally locked exoplanets such as Trappist-1 e \citep{Sergeev2020, Sergeev2022}.

The right-hand plot in Figure \ref{fig:casestudy_ts_olr_cld} shows the map of total cloud cover. Clouds are concentrated in the substellar region and immediately to the east of it, and are present generally as tropospheric clouds there, present in the convective region below 300mbar and in the convective inhibition layer right at the surface. There are fewer nightside clouds, suggesting that clouds have a net radiative cooling effect. However, this is limited by the fact that most of the longwave emission happens higher up in the atmosphere than the clouds, which are present mostly in the lower troposphere. Despite these high cloud fractions, they do not substantially contribute to the planetary albedo, as discussed further in Section \ref{sec:r3_rs}. This is because there is substantial absorption higher in the atmosphere from H$_2$ as well as CO$_2$ and CH$_4$, so relatively little flux reaches the tropospheric clouds.

\subsection{Dynamical Structure}

In many aspects the dynamical structure matches what we would expect from a tidally-locked terrestrial exoplanet \citep{Showman2013, Pierrehumbert2019}. `Terrestrial' in this context effectively means a shallow atmosphere, so this is a regime we are in despite the Hycean nature we are investigating for K2-18b. The key parameters of the planet, notably its radius ($R_p = 2.6 R_\oplus$, significantly larger than terrestrial planets), its slow rotation rate ($\Omega =2.2\times 10^{-6}$ rad s$^{-1}$), alongside its relatively low instellation and H$_2$ dominated atmosphere, all contribute to placing it roughly in a `slow rotator' regime. Two key parameters to help us further understand the dynamical regime are the dimensionless equatorial Rossby deformation length $L_{Ro}$ and the dimensionless Rhines length $L_{Rh}$. Following standard definitions, e.g. \citet{Haqq-Misra2018}, 

\begin{equation}
    L_{Ro} = \sqrt{\frac{NH}{2\Omega R_p}} \label{eq:l_ro}
\end{equation}
\begin{equation}
    L_{Rh} = \pi \sqrt{\frac{U}{2\Omega R_p}} \label{eq:l_rh}
\end{equation}

\begin{figure*}
    \centering
    \includegraphics[width=0.95\textwidth]{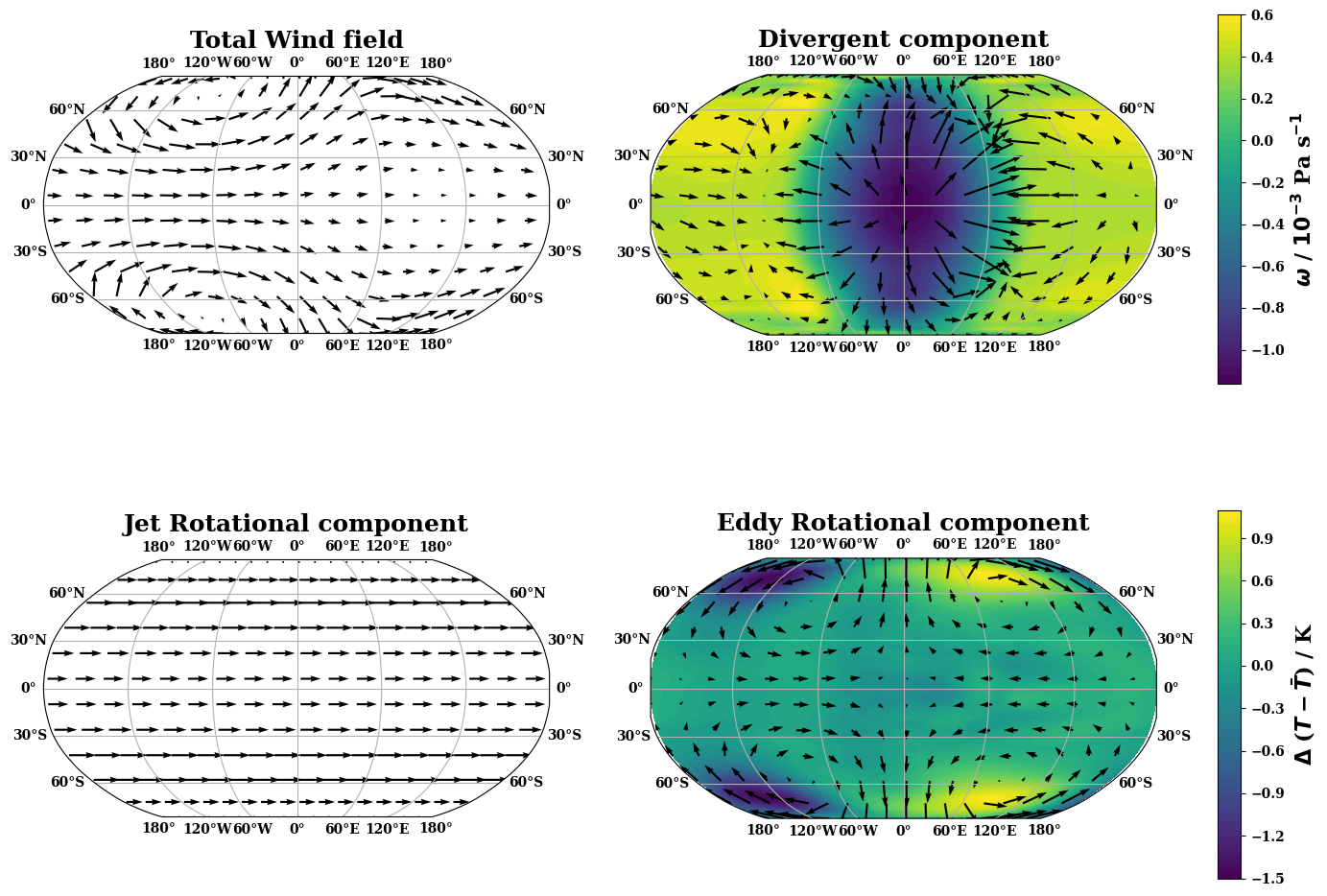}
    \caption{Wind fields in our imposed TOA $A_b=0.6$ case. Top left: horizontal wind field at 250 hPa. Top right: Divergent component of the circulation, and the vertical pressure velocity $\omega$ at 100 hPa. Bottom left: Zonal-mean rotational (Jet) component Bottom right: Eddy rotational component, and deviations of temperature from the zonal mean at 600 hPa. The eastward nature of the overall winds can be seen, although it is strongest moving from the substellar  to the antistellar point. The wind's jet rotational component makes up the largest single component of the wind and shows twin high-latitude jets. The eddy rotational component is also significant and shows equatorial Rossby waves. In the western hemisphere these increase the eastward wind magnitude at the equator and reduce it at the pole, whereas the opposite is true in the eastern hemisphere. Compared to the rotational component, the divergent circulation is small and shows roughly strong divergence at the substellar point motion, although this is compensated by a returning circulation at lower pressures. Despite its small magnitude, the divergent circulation is still responsible for the majority of the net dayside-nightside heat transport. The jet and eddy rotational components carry a significant amount of heat but largely cancel each other out.}
    \label{fig:casestudy_helmholtz}
\end{figure*}

H is the scale height, N is the Brunt-Vaisala frequency, and U is the zonal wind; these values are calculated by a mass weighted tropospheric average between 0.1 and 1 bar. Following this, we find $L_{Ro} \approx 1.5$ and $L_{Rh} \approx 1.9$. This places us squarely in the slow rotator regime \citep{Haqq-Misra2018}. This translates to equatorial Rossby waves, an eastwards shift of the maximum surface heating, a broad equatorial jet, a strong upper-atmosphere super-rotation, and a general dayside-nightside overturning circulation, as shown in Figure \ref{fig:casestudy_helmholtz}. 

We start by establishing the presence of two anticyclones to the east of the substellar point and two cyclones to the west of it. This is shown by the eddy rotational wind component in the bottom right of Figure \ref{fig:casestudy_helmholtz}, where we perform a Helmholtz decomposition of the wind at the 200 hPa level. This pattern corresponds to the westwards propagating equatorial Rossby wave \citep{Sergeev2020}, and is matched by middle-tropospheric cold spots in the cyclones and hot spots in the anticyclones, also shown in the figure as deviations from the zonal mean. Once the jet zonal winds are taken into account, this leads to western hemisphere wind gyres in the coldest regions of the planet (Fig. \ref{fig:casestudy_helmholtz}, upper-left). This dynamical structure is similar to that found in - for example - the 'MassFlux' Trappist-1 e case \citep{Sergeev2020} and the Rocke-3D Ben 1 case \citep{Turbet2020}; it is in opposition to the eastern hemisphere cyclonic gyres which commonly mark 'fast rotator' cases, such as in the ExoCAM, LMD-G, and UM Trappist-1 e Ben 1 cases \citep{Turbet2022} and the 'Adjust' and 'NoCnvPm' Trappist-1 e cases \citep{Sergeev2020}. These cases are also marked by cold spots at these gyres and corresponding hot spots in western hemisphere, representing an extra-tropical Rossby wave pattern \citep{Carone2015}.  %The gyres are largely barotropic and the temperature contours largely follow the eddy geopotential. 

A key element of the dynamical structure is our zonal wind structure, the zonal-average of which is shown in the right panel of Figure \ref{fig:casestudy_uzonal_divsu}. It shows a general super-rotation of the atmosphere: there are no westerly winds anywhere. Although the atmospheric state was initialised at rest, the surface drag provides a net torque and leads to this net angular momentum. There are mid-latitude zonal jets in the stratosphere reaching speeds of $\approx 140$ ms$^{-1}$, and a very broad equatorial tropospheric super-rotating jet as has often been the case in simulations of tidally-locked exoplanets \citep[e.g.][]{Charnay2021, Sergeev2022, Turbet2022}. Here the slow rotation speed of the planet means that the equatorial jet is not constrained to low latitudes and extends across much of the planet.%Showman2013, Pierrehumbert2019, Hammond2020, 

\begin{figure*}
    \centering
    \includegraphics[width=0.95\textwidth]{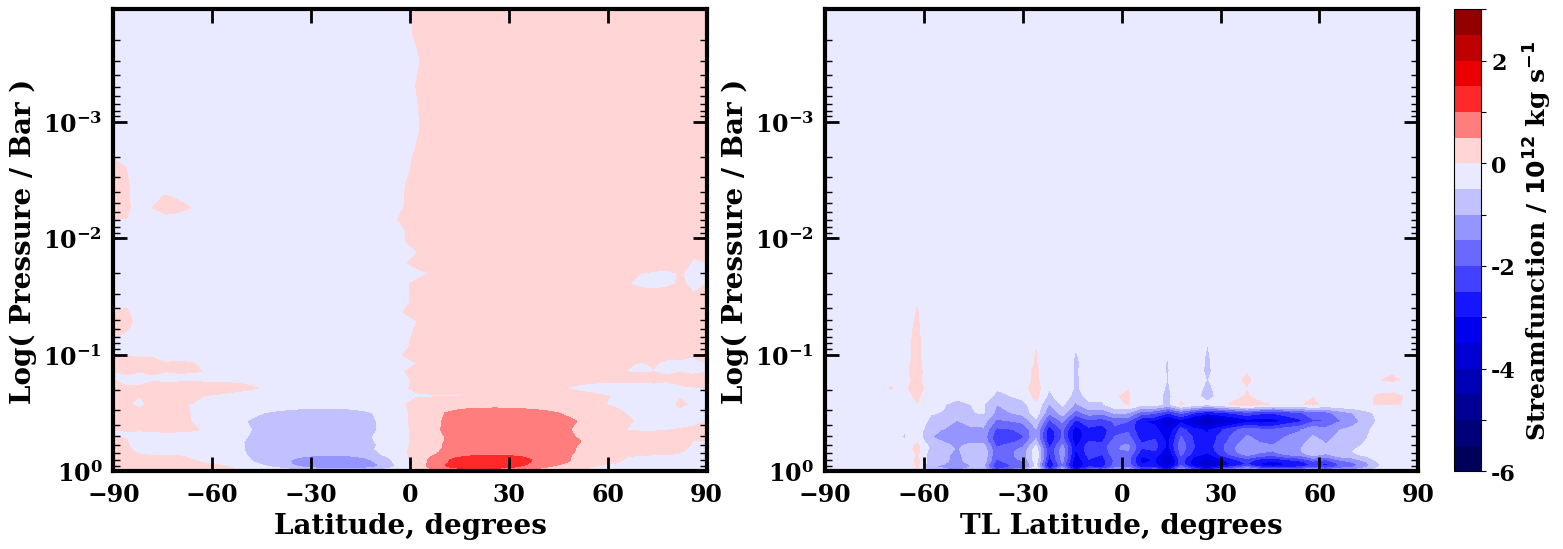}
    \caption{Dynamical features in our imposed TOA $A_b=0.6$ case. Left: Mean meridional mass streamfunction in the standard latitude-longitude coordinates, in units of $kg$ $s^{-1}$. Right: Mean meridional mass streamfunction in Tidally-locked latitude-longitude coordinates, in units of $kg$ $s^{-1}$. A TL latitude of 90 (-90) $\degree$ corresponds to the substellar (antistellar) point. Average circulation is anticlockwise around negative values and clockwise around positive values. We see both equator-pole and a more general dayside-nightside overturning circulation. The dayside-nightside overturning circulation is stronger than the pole-equator overturning circulation, but of the same order of magnitude.}% Of course, the two are not independent - the equator-pole circulation may only be a slice in the dayside-nightside circulation and there may not be two independent circulations.}
    \label{fig:casestudy_msfs}
\end{figure*}

\begin{figure*}
    \centering
    \includegraphics[width=0.95\textwidth]{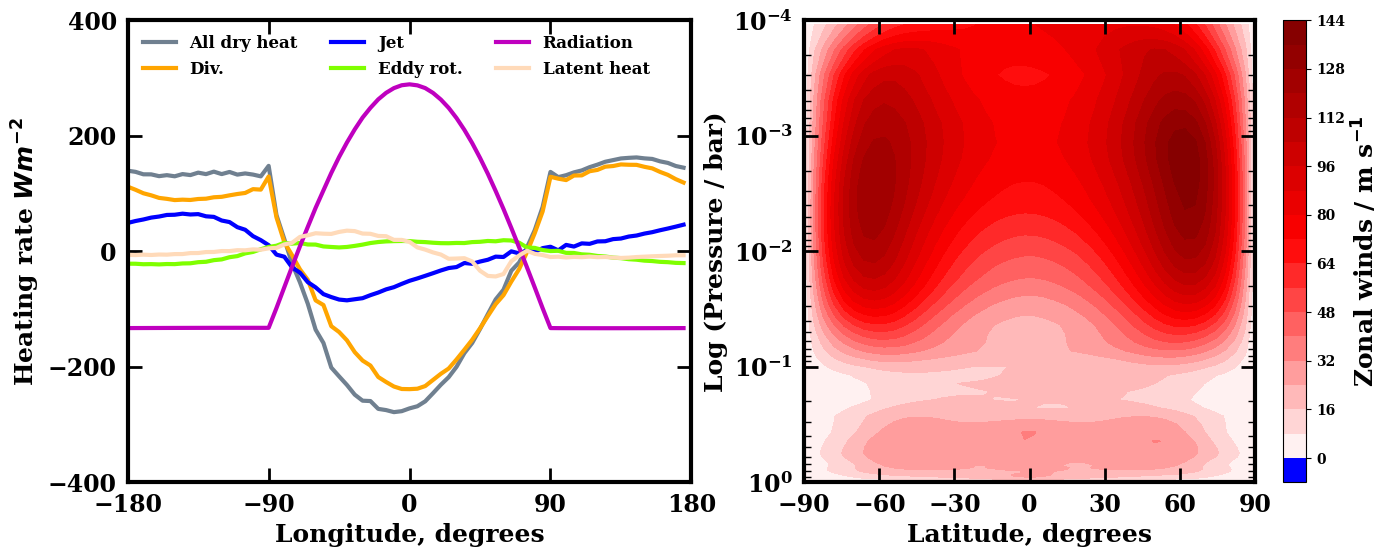}
    \caption{Dynamical features in our imposed TOA $A_b=0.6$ case. Left: The contributions of stellar heating and atmospheric circulation to the large-scale heating patterns, as shown for each longitude slice, averaging over all altitudes and latitudes. The divergent circulation makes up the majority of the day-night heat flux despite its much smaller magnitude. The jet and eddy rotational components carry a significant amount of heat but largely cancel each other out, and the latent heat flux also carries some energy, but the low H$_2$O abundances in the free troposphere limit its magnitude. There are some small residuals (not shown) left from regridding approximations. Right: The zonally averaged zonal wind \textit{u}. The atmosphere as a whole is super-rotating with no westerly winds anywhere, which is possible given the presence of surface drag. We see twin stratospheric jets at high latitudes, with one set in the troposphere and then stronger stratospheric twin jets. There is a single broad equatorial jet in the troposphere.}
    \label{fig:casestudy_uzonal_divsu}
\end{figure*}

We expect some degree of overturning circulation to transport heat away from the substellar region. In particular, in our slow-rotating regime, we anticipate a general overturning circulation with the bulk atmosphere being heated and rising on the dayside with corresponding subsidence on the nightside. We can see this overturning circulation by plotting the meridional mass streamfunctions in Figure \ref{fig:casestudy_msfs}. The meridional mass streamfunction represents a meridional mass flux between the top and the bottom of the atmosphere \citep{Innes2022}. Circulation is on average clockwise (anticlockwise) around positive (negative) values. There is both a single broad Hadley cell between the Equator and Pole, and a single dayside-nightside overturning circulation. The dayside-nightside circulation is a few times stronger than the equator-pole one, but both are of the same order of magnitude. This is a similar result to other GCM simulations of K2-18b in a mini-Neptune state \citep{Innes2022, Barrier2025}, and show that details such as these are strongly affected by the planet radius and rotation rate, which control the relative influence of rotation on the system.

We can also see this overturning more directly in the upper right panel of Figure \ref{fig:casestudy_helmholtz} which shows the divergent component of the circulation as well as the vertical pressure velocities at the 100 mbar level. This divergent circulation is not strong in absolute terms, especially compared to other tidally-locked GCM simulations with mid-latitude jets (e.g. ROCKE-3D Hab 1, \citet{Sergeev2022}). It does, however, perform the bulk of the dayside - nightside heat transport, as shown in the left panel of Figure \ref{fig:casestudy_uzonal_divsu}. Following the same technique as e.g. \citet{Sergeev2020} and \citet{Hammond2021}, we calculate longitude-averaged heating from radiation and the divergence of moist static energy  $\nabla \cdot h\underline{u}$, splitting $h$ into dry static energy $s$ and the water vapour component $Lq$, and $\underline{u}$ into its different wind components. The latent heat flux is relatively small, as there is little water in the free troposphere. It is the divergent component which is the main source of dayside-nightside heat flux. The jet and eddy rotational components still transport significant amounts but also mostly cancel each other out. This dominance of the divergent component can largely be explained by the planet being in a WTG regime following the reasoning in \citet{Hammond2021}. Expanding $\nabla \cdot s\underline{u} = s \nabla \cdot \underline{u} + \underline{u} \cdot \nabla s$, the irrotational component of the wind will have no divergence, eliminating the first component, whilst the second component is small due to the WTG regime. In contrast, the divergent portion of the wind will have a significant $s \nabla \cdot \underline{u}$ term and so will provide the majority of the heat transport. Here there are still some tropospheric temperature differences, and so the rotational components carry more heat than e.g. the mini-Neptune case in \citet{Barrier2025}, but the divergent component is still dominant.

\subsection{Convection and the substellar region}

A particularly pertinent topic with possible Hycean planets, and hydrogen-dominated atmospheres with a liquid water boundary in general, is the presence of convective inhibition and the impact this might have on the climate state, convective patterns, and runaway greenhouse limit \citep{Leconte2017, Innes2023, Leconte2024, Habib2024b, Seeley2025}. Modelling on this topic has so far involved either 1D models, or high-resolution Convection Permitting Models (CPMs), and not a global 3D model. To compare our results, we first look at where convection is active and what its global impact looks like. We then take a look at the substellar region specifically to compare our results to high resolution work. In this section, we define the substellar region as being anywhere within 50 \degree E/W or 30 \degree N/S of the substellar point, which is a somewhat arbitrary definition but ensures that all the region we examine has significant convection occuring.

\begin{figure*}
    \centering
    \includegraphics[width=0.95\textwidth]{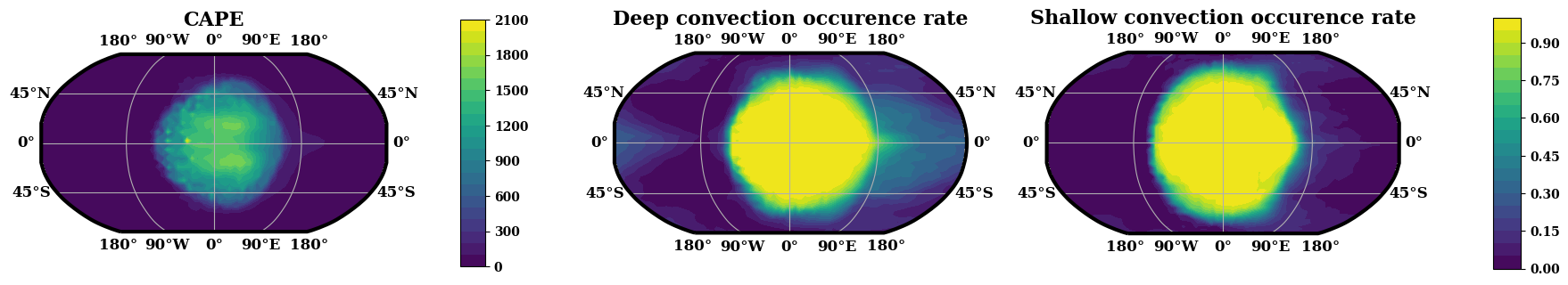}
    \caption{In our imposed TOA $A_b=0.6$ case, the convective available potential energy (CAPE), in J kg$^{-1}$. Middle: the occurrence rate of deep convection in the column. Right: the occurrence rate of shallow convection. Deep and shallow convection are both omnipresent in the substellar region, and present to a degree outside it, especially in the equatorial band. CAPE is significant in the substellar region, and increases towards the substellar point.}
    \label{fig:casestudy_capefreqzmsh}
\end{figure*}

We begin with the spatial distribution of CAPE and the occurence rates of deep and shallow convection in Figure \ref{fig:casestudy_capefreqzmsh}. There is near constant shallow and deep convection in the substellar region, with shallow convection - which does not have a minimum CAPE threshold for triggering - covering a slightly greater area. There are also significant amounts of convection happening outside this substellar region, especially in the equatorial band. This happens higher up in the atmosphere: the near-surface regions are generally not convectively unstable but higher up, where the pressure-temperature profiles are more globally uniform, they can be. This image of substellar-point centred convection is supported by the CAPE profile. CAPE is concentrated in the substellar region, rising towards the substellar point, and is very small outside this. 

\begin{figure*}
    \centering
    \includegraphics[width=0.95\textwidth]{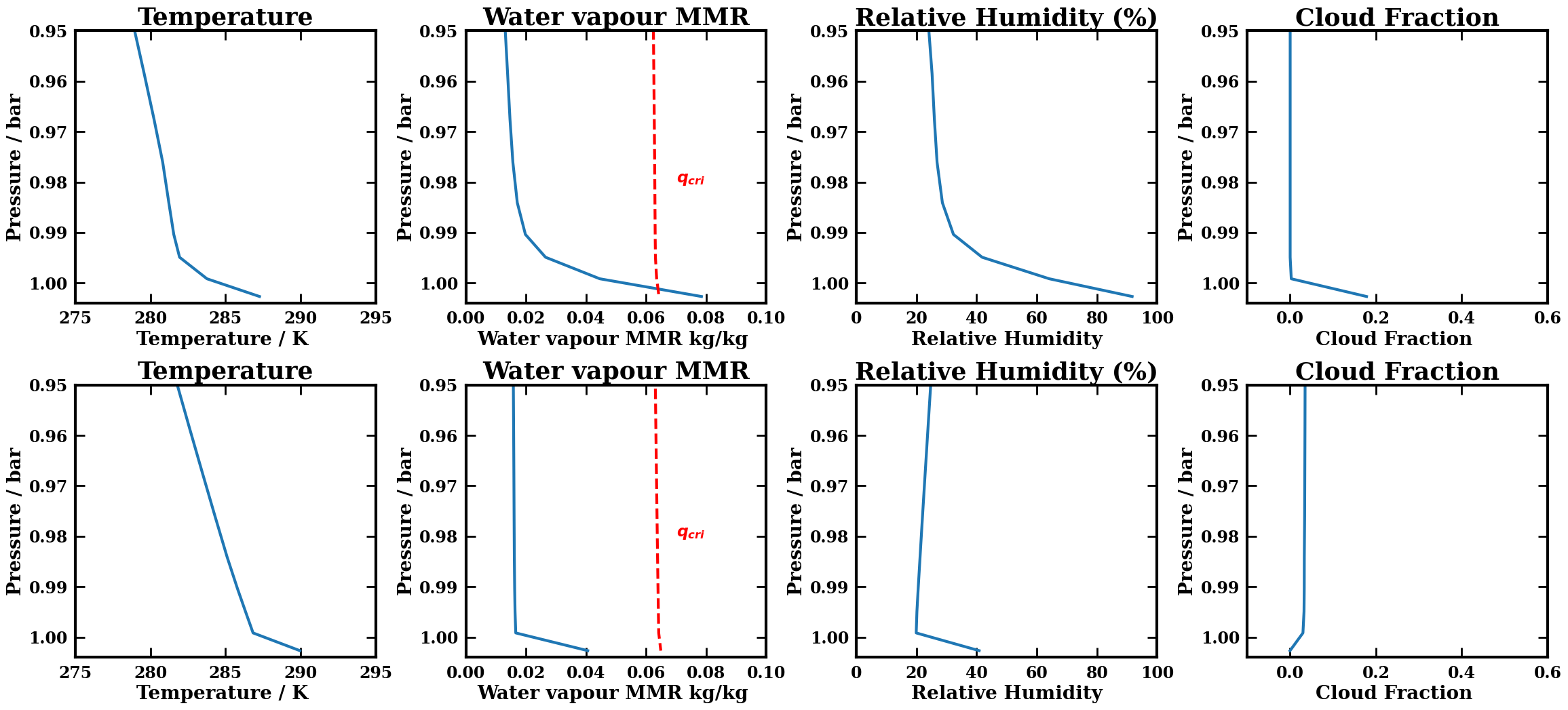}
    \caption{Profiles of key parameters in the near-surface inhibition zone (top row) and the near-surface convective zone (bottom row) in our imposed TOA $A_b=0.6$ case. Left: Pressure-Temperature profile. Middle Left: Pressure- Water MMR profile, with the critical water MMR $q_{cri}$ also shown. Middle Right: Relative Humidity profile. Right: Cloud profile. The behaviour of our convective zone closely matches that expected from high-resolution simulations, but an alternative situation is also seen. In this zone with convective starting from the surface, the surface layer has a notably higher $q$ than the layers directly above it due to surface ocean evaporation into it. This compositional stratification also corresponds to a slight increase in the lapse rate - if not for this increased lapse rate the surface layer would be stable to convection. Relative humidity also increases near the surface, but crucially stays low enough that $q<q_{cri}$. No surface level clouds develop.}
    \label{fig:casestudy_mcinomci}
\end{figure*}

To establish the existence of convective inhibition on a granular level, we look for the presence of regions whose near-surface lapse rates are greater than the moist adiabat, but where $q$ is greater than the critical water MMR $q_{cri}$ above which convective inhibition develops. In our simulations, these regions are indeed present.
%, as we suspected from Figure \ref{fig:casestudy_cconv}. 
A key question we seek to answer is what these regions look like, and in particular how their  emergent properties as we find in our 3D global simulations compare to those found in high resolution CPM simulations, notably those of \citet{Seeley2025}. The key properties of our convective-inhibition zones are shown in the top row of Figure \ref{fig:casestudy_mcinomci}, where we show the temperature, moisture, relative humidity and cloud profiles as averaged across the inhibition region.

The top-left panel of Figure \ref{fig:casestudy_mcinomci} shows, as expected, steep temperature gradients in the near-surface regions - which here is only present in the few layers above the surface, with a net thickness of about 1km. The lapse rate in the bottom layer is about 20 K / km, which is roughly in line with the expectations of Figure 5 of \citet{Seeley2025} for a $\mu_d=3.16$, $T_s=290$ system. As expected, the inhibition layer is almost completely saturated (RH $\approx 0.9$) and cloudy. Above this inhibition layer, moist convection does develop. We see a strong reduction in the lapse rate, a flatter $q$ profile, and the subsaturation associated with moist convection. This shows that our GCM parametrisations, including the convection and turbulence parametrisations, are capable of reproducing the main qualitative and quantitative predictions from CPM simulations. It also shows that moist convective inhibition does apply in a global context, and that large-scale zonal winds and the overturning circulation does not seem to affect the presence of moist convective inhibition as was theorised to be possible in \citet{Innes2023}.

However, we do find another very interesting result: moist convective inhibition does not seem to be the only possible state possible. In fact, a large portion of the substellar region sees fairly standard moist convection, which leads to the profiles in the bottom row of Figure \ref{fig:casestudy_mcinomci}. The key difference in this case is that moist convection reaches down the surface and efficiently removes moisture from those layers. This keeps the surface layer $q$ low enough - for $T=290$ K and $\mu_d=3.162$, $q_{cri}=0.065$ and we see from the bottom, second panel from left of Figure \ref{fig:casestudy_mcinomci} that the average $q$ is below this. The subsaturation of the surface layer is crucial to this: if it were saturated $q$ would be high enough to inhibit moist convection. Instead moist convection develops, although the convection is strictly speaking dry until the updrafts reach the lifting condensation level (LCL) and condensation happens.

How these two climate states co-exist deserves further study. Long term time-averaged values of near-surface temperature and moisture show geographically smoothly varying averages and no evidence of two different states, indicating that the regions of the surface shift back and forth between the two states. Time series of surface conditions in single grid cells indicate that the transition in and out of convective inhibition happens on timescales of tens of days, which is longer than the variability of $\mathcal{O}(days)$ found in \citet{Seeley2025}.

It seems that normal time variability, driven by large scale dynamical motions, is enough to move sections of the surface from one regime into the other. How exactly these transitions happen in these conditions, and why we do not see the day-scale variability seen in some high-resolution models, deserves further investigation. It is likely to depend on the precise properties of the local and global environment, and it is also possible that our mass-flux convective scheme is not capturing some physical process essential for the short-term variability.  

\section{Results II - Effect of TOA albedo}
\label{sec:r2_ab}

We now present of the rest of the imposed TOA Albedo simulations. We first focus on the runaway greenhouse constraints found: we are particularly interested in the threshold at which a stable Hycean climate stops being possible, and so what set of albedos would be required for a Hycean K2-18 b to exist. Next, we explore how the rest of the $P_s=1$ bar cases change from the reference $A_b=0.6$ case, before discussing the climate states of the $P_s=5$ bar cases. The simulations carried out are all listed in Table \ref{tbl:ab_runs}.

\subsection{Liquid surface water stability threshold}

The runaway greenhouse effect \citep{Komabayasi1967, Ingersoll1969, Nakajima1992} is a well-known phenomenon in planetary science where an increase in surface temperature leads to an increased amount of water vapour in the atmosphere, leading to an increased total atmospheric optical depth and a larger greenhouse effect. As the photosphere gradually rises away from the surface, then at some point the surface greenhouse heating induced will outweigh the cooling from increased thermal radiation outward. An unstable climate state is reached where the surface temperature continuously increases without increasing the OLR, which reaches a limiting value. This state continues until either the surface liquid reservoir is completely evaporated (which would take an extraordinarily long time on a Hycean world given the size of the H$_2$O layer) , or until sections of the atmosphere are hot enough to emit in optical or near-infrared bands where the H$_2$O opacity is lower.

In our runs we can recognise the start of the runaway greenhouse threshold through a steady increase in surface temperature on the dayside coupled with a constant OLR. Eventually the GCM becomes unable to handle this transitional state and the run crashes. The rate of the temperature increase is reasonably slow, at least compared to other GCM simulations of a runaway greenhouse developing in Earth-like conditions \citep{Chaverot2023}. This can be explained by the lower instellation in our cases and the photosphere already being located in the free troposphere, so that there is less stellar energy entering the near-surface layers and contributing to the heating. Notably, there is still relatively little change in the total surface pressure and the total water column mass. This is due to the large amount of heat required to evaporate water, and leads to considerable additional subsaturation in the near-surface regions, even in regions where moist convection is inhibited. When discussing the results in more detail below, in the cases where no stable climate was reached, we use the values from just before the crashes. Some of these atmospheres are correspondingly not in equilibrium and their results should be interpreted as such.

\begin{figure}
    \centering
    \includegraphics[width=0.48\textwidth]{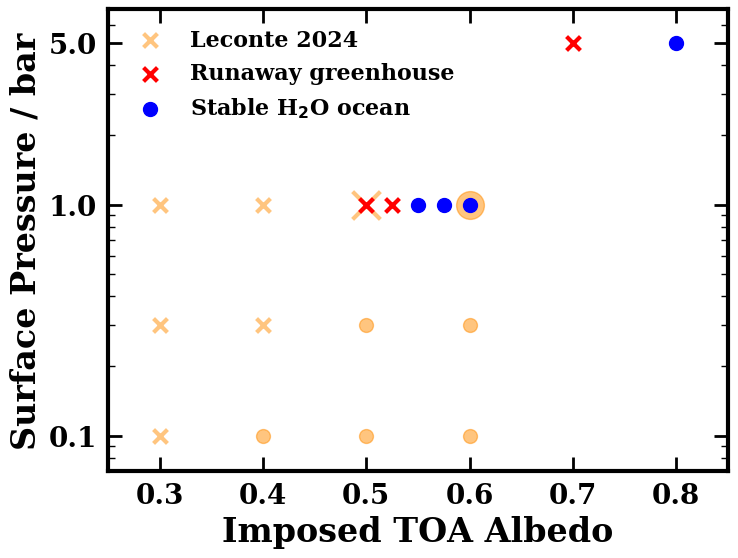}
    \caption{Constraints on the stability of a Hycean K2-18b as calculated by imposing a TOA albedo. The red crosses denote cases undergoing a runaway greenhouse transition, with the blue circles corresponding to states with a stable Hycean climate. Overplotted are the results of \citet{Leconte2024}, who performed similar work using a 1D climate model. For a 1 bar atmosphere, the runaway greenhouse threshold is at an albedo of between 0.525 and 0.55. For a 5 bar case, this same threshold is between 0.7 and 0.8 - similar to the albedo of Venus.}
    \label{fig:r2ab_mainresult}
\end{figure}

Our albedo constraints are presented in Figure \ref{fig:r2ab_mainresult}, which shows which GCM runs led to stable climate states and which did not. Overplotted are the values from Figure 7 of \citet{Leconte2024}. We run simulations at surface pressures of 1 and 5 bar instead of 0.1 and 1 bar, and our results are from 3D global circulation models whereas theirs are from a 1D radiative-convective model, informed by high resolution CPMs. We also use slightly different atmospheric compositions: \citet{Leconte2024} used CH$_4$ and CO$_2$ volume mixing ratios of 10$^{-2}$, whereas we use the likeliest abundances from \citet{Madhusudhan2023b} (CH$_4$ VMR: 1.82$\times 10^{-2}$, CO$_2$ VMR: 8.13 $\times 10^{-3}$)which lead to a slightly greater greenhouse effect. Despite this our results are consistent with each other.

We find in this case that the threshold for the stable presence of liquid water, also known as the Inner Habitable Zone or IHZ, lies between albedos of 0.525 and 0.55 for a 1 bar Hycean K2-18b and between 0.7 and 0.8 for a 5 bar Hycean. This matches the results of \citet{Leconte2024} who found that the boundary was located between albedos of 0.5 and 0.6 for a 1 bar atmosphere. 
Cases with an albedo higher than the critical albedo continue to possess liquid water oceans. Those with a lower albedo, on the other hand, are unstable and undergoing a runaway greenhouse transition.

We note that the agreement between 3D and 1D modelling is not necessarily guaranteed: geographic variations in convection can plausibly alter the location of this threshold depending on how efficiently heat, moisture, and mass is transported from the substellar region to the rest of the planet. As the near-surface moisture fraction is exponentially dependent on temperature, the surface variations in temperature we saw (e.g. in Figure \ref{fig:casestudy_ts_olr_cld}) mean that the substellar regions reaches a runaway greenhouse state first (in the sense that the extra surface temperatures there lead to increased water vapour and so increased IR optical depth and so on...). This extra heat is not radiated away and instead contributes to a global warming of the near-surface regions. The runaway greenhouse is thus an inherently 3D effect in its initialisation and development. In this case, however, the minimal temperature differences in the free troposphere where the photosphere is located mean that the substellar region P-T profiles closely align with what a 1D model would give, and so the results are similar.

\subsection{Cases with 1 bar surface pressure}
 
\begin{figure*}
    \centering
    \includegraphics[width=0.95\textwidth]{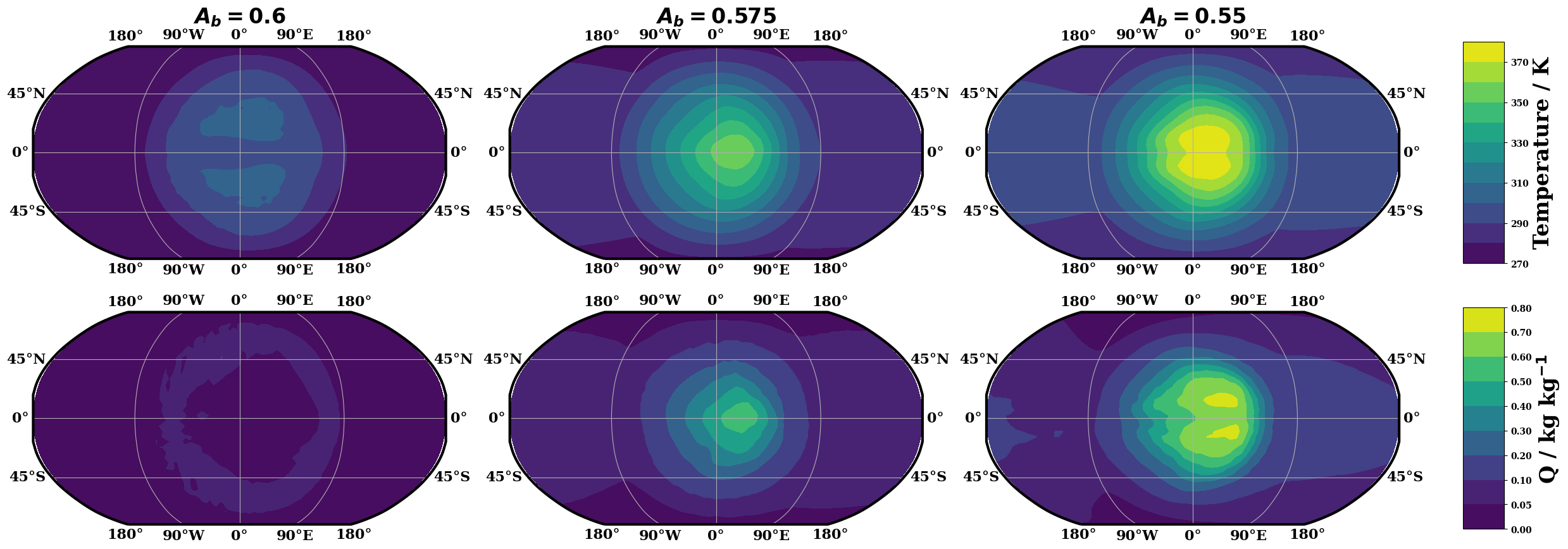}
    \caption{The surface temperatures and moisture mass fraction for the 1 bar cases with imposed TOA albedos. A steady trend in increased T$_s$ and $q_s$ is seen with increased instellation. $q_s$ also rises to very high values - it lags behind the temperature somewhat, leading to subsaturation, because of the energy needed to evaporate that amount of water. Surface temperature and moisture gradients both increase as instellation increases.}
    \label{fig:r2ab_TQSs}
\end{figure*}

We now take a closer look at 1 bar surface pressure cases, and present an overview of the key features of the climate states and trends in them. We start by showing the surface temperatures and water mass fractions of the cases in Figure \ref{fig:r2ab_TQSs}. As expected, we see a steady trend in increased temperature and water mixing ratio with increased instellation. In the hottest cases (not shown), the surface temperature is in fact greater than the boiling point of water at 1 bar, a reflection of the disequilibrium present. The magnitude of the horizontal temperature contrasts also increases significantly with instellation from about 30K in the $A_b=0.6$ case to about 80K in the $A_b=0.55$ case, a reflection of the greater convective inhibition that massively increases dayside surface temperatures without much resulting impact on the free tropospheric temperatures.

These surface temperature and moisture values can be compared to those in Figure 3 of \citet{Kuzucan2025}, who also investigated the IHZ for a 1 bar H$_2$-rich atmosphere (but with some key differences in their model: they use the bulk parameters of Earth, a purely H$_2$-H$_2$O atmosphere, and assume the planet is a fast rotator). We see that our planetary averages are similar, but that the peak temperatures and moistures are relatively higher in our case. This is likely because our tidally locked configuration means a relatively much higher instellation in the substellar region, and also potentially due to the inclusion of convective inhibition in our simulations.

The variations in temperature and moisture with height follow expected patterns. There is a general increase in upper tropospheric and stratospheric temperatures of 20K between the 0.6 and 0.5 albedo cases (with the PT profile for the former shown in Figure \ref{fig:casestudy_PTs}), a consequence of the necessity for the OLR to increase and match the incoming stellar radiation. The lower tropospheric temperatures increase by a corresponding amount, and the temperature difference between the two is controlled by the moist adiabatic gradient. 

Conversely, the temperature at the substellar point does increase very sharply as a result of a deeper and steeper super-adiabatic inhibition zone. This can be explained by considering that the top of the inhibition layer is always at roughly the temperature where a saturated medium is just at the Guillot threshold. If this temperature is reached higher in the atmosphere, the inhibition zone (stretching to the surface) is deeper, and the higher $q$ gradients lead to a higher lapse rate, as found in \citet{Seeley2025}. The moisture profiles follow the expected pattern - constant above the tropopause, then steadily increasing throughout the troposphere (bar a small region of dry shallow convection in some cases) before it spikes in the near-surface inhibition zone.

In the hotter $A_b<0.6$ cases, interestingly, we see a departure from the saturated surface layer seen in cooler cases. There is still an inhibition zone with very high lapse rates, and the water mass mixing ratio (MMR) and saturation both sharply increase towards the surface, but the layer remains subsaturated (relative humidity rises to around 60\%) and there is at most only a slight increase in cloud cover. In this case convection is stopped by compositional stratification and not moist convective inhibition specifically, showing that both mechanisms can lead to a near-surface inhibition layer.

\begin{figure*}
    \centering
    \includegraphics[width=0.95\textwidth]{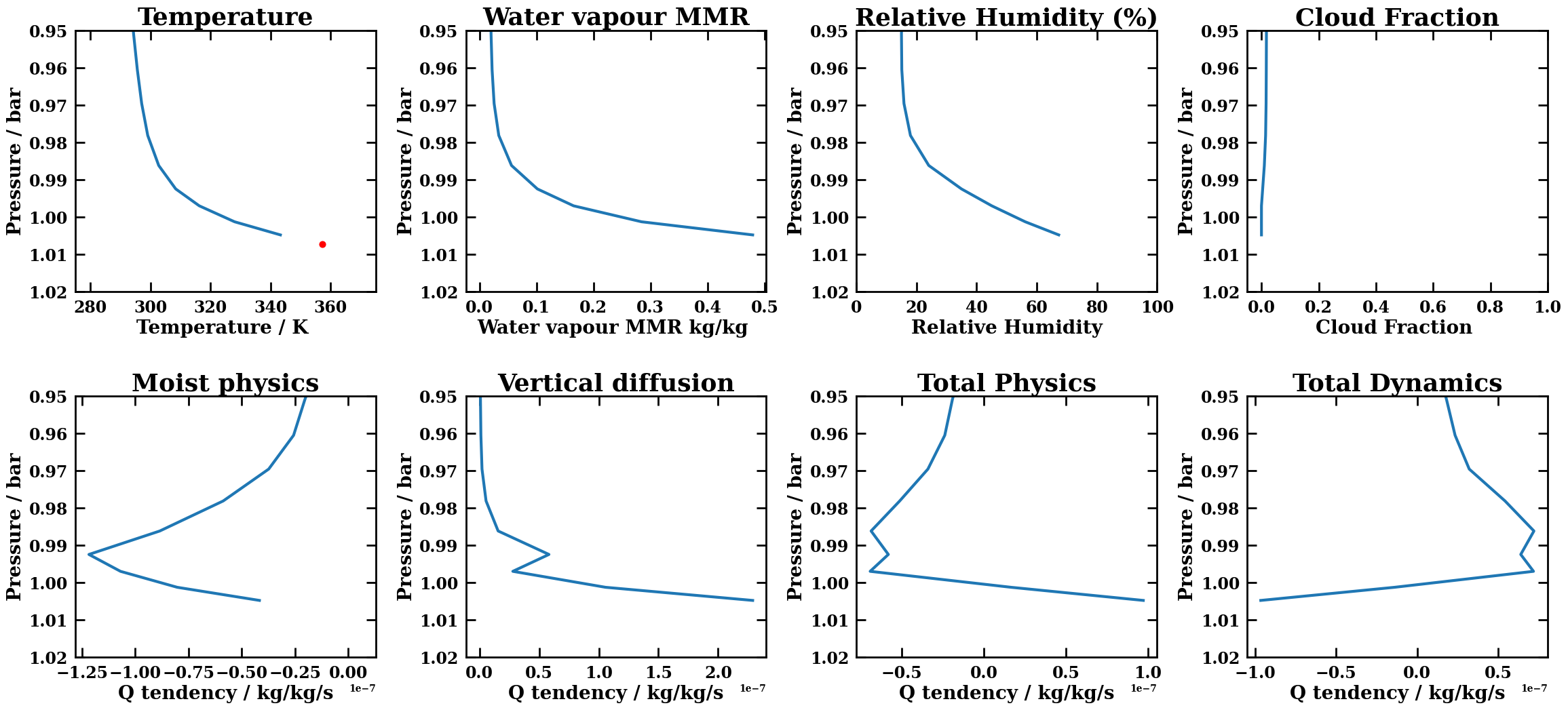}
    \caption{The profiles of key variables, including the main contributors to the water budget, to the substellar zone of our imposed TOA $A_b=0.575$ case.  Left: Pressure-Temperature profile, with the sea surface temperature also shown as the red dot. Middle Left: Pressure- Water MMR profile. Middle Right: Relative Humidity profile. Right: Cloud profile. In marked contrast to the cooler case shown in Figure \ref{fig:casestudy_mcinomci}, the inhibition zone is subsaturated, even at the surface, although there are sharp temperature and moisture gradients. An analysis of the water budget of the near surface region reveals that evaporation into the surface layer (accounted for in the vertical diffusion profile) is balanced by the convection, vertical diffusion upwards, and the drying caused by large scale dynamical motions.}
    \label{fig:r2_ab_1b_582w_surf}
\end{figure*}

Why this subsaturated inhibition layer develops is a good question. It appears to be the case that in cases like these where the surface saturation H$_2$O VMR is very high, the sheer volume of water that needs to evaporate to keep the surface saturated means that other processes (largely turbulence and large scale dynamical motions) remove water on a fast enough timescale that full saturation is not reached. Another important factor which characterises these cases is that the top of the inhibition zone in these cases is quite dry, with a humidity of 10-30\%. The profiles of key quantities through this near surface zone is shown alongside the contribution of various model sections to the water budget in Figure \ref{fig:r2_ab_1b_582w_surf} for the $A_b=0.575$ case. 

The exact behaviour of the inhibition zone in these conditions is complex and deserves much more study with different models - it is not completely excluded that some limitation in our GCM or our convection scheme is leading to this subsaturated zone. We have run a number of tests tuning various variables to see their effects, the results of which we briefly describe here. In the first test, we increased the magnitude of the surface heat and moisture fluxes by a factor of 10. This counterintuitively leads to a collapse of the surface inhibition zone and a subsequent reduction in surface temperature. Naively, it might have been expected to lead to saturation and hotter temperatures, but the increased heat flux seems to have created an unstable PBL and so led to much greater PBL transport of heat and moisture, destroying the inhibition zone.

In the second test, an order of magnitude increase in $K_{zz,min}$ leads to a slightly more moist upper inhibition zone but otherwise has a small effect. In our third and fourth tests, we change the vertical extent of the sub-cloud heating and drying in our convective parametrisation. This is normally spread over the start level of convection and the two levels below that \citep{Barrier2025}, which could plausibly be unphysically removing moisture from the inhibition zone. However, changing the drying so it either happens completely in the convection start layer (test 3) or all the way to the surface (test 4) has minimal changes to the inhibition layer: although the convective drying tendencies do change, the final effect is still a subsaturated surface layer.

The dynamical structure of the runs stays broadly similar throughout most of our runs, with equatorial Rossby lobes, mid-latitude stratospheric zonal jets, and divergent circulation at the stellar point. The  magnitude of the overturning circulation increases with increasing instellation, a reflection of the greater dayside-nightside energy transport, and the strength of the winds also increases with the stratospheric jets reaching $\approx 200$ ms$^{-1}$ . Some differences do emerge, particularly with the zonal winds. As instellation increases, instead of a single broad equatorial jet, we see mid-latitude jets, with the axial angular momentum also peaking in the mid-latitudes. Additionally, in the developing stages of its runaway greenhouse, the the 0.5 albedo case (685 Wm$^{-2}$) is overall in a fast rotator regime, as diagnosed by its extra-tropical Rossby waves and corresponding Eastern hemisphere cold spot. 
These changes are likely linked to the increased instellation and so greater equatorial-pole temperature differences. This greater baroclinicity drives stronger extratropical Rossby waves and  cyclostrophic zonal jets, following the picture in \citet{Sergeev2020}. In support of this view, we do observe a decline in $L_{Ro}$ from 1.5 in the $A_b=0.6$ case to about 1.2 in the $A_b=0.55$ case and to about 1 in the $A_b=0.5$ case. There are also layers of the troposphere with Rossby numbers lower than the tropospheric average, further favouring the formation of extratropical Rossby waves there. Care must be taken to not interpret these results too far, particularly as the $A_b=0.5$ climate state is in the early stages of a runaway greenhouse. However, it does serve as a reminder of the importance of instellation on the climate state and how general, planet-averaged scaling arguments such as Equations \ref{eq:l_ro} and \ref{eq:l_rh} will not always give the complete dynamical picture, particularly with regard to the formation of zonal jets.

\subsection{Cases with a 5 bar surface pressure}

So far we have focused on climate states with a 1 bar atmosphere. This is an arbitrary restriction: canonical Hycean atmospheres can in theory go up to 100 bar \citep{Madhusudhan2021}, as long as temperatures are low enough for there to be liquid water at the surface. In our case we use a 5 bar atmosphere as our test bed to see the effects of increased surface pressure on the climate state. This is also a somewhat arbitrary choice, but - as outlined in Section \ref{sec:methods} - it is justified by the spin-up times required for deep albedos, by the motivation of finding plausible albedos for stable Hycean climates, and to allow comparison with previous work focusing on constraints on shallower atmospheres. Given that they share the same atmospheric composition, planet size, rotation rate, and surface boundary conditions, the 5 bar case unsurprisingly has many similarities to the 1 bar case. There are, however, some important differences due to the increased dry mass of the atmosphere.
Due to the increased optical depth to the surface, we are no longer in a terrestrial regime. However, enough stellar flux reaches the surface that we are not quite yet in a deep atmosphere regime where the presence of a surface has a negligible effect on the climate.

\begin{figure*}
    \centering
    \includegraphics[width=0.95\textwidth]{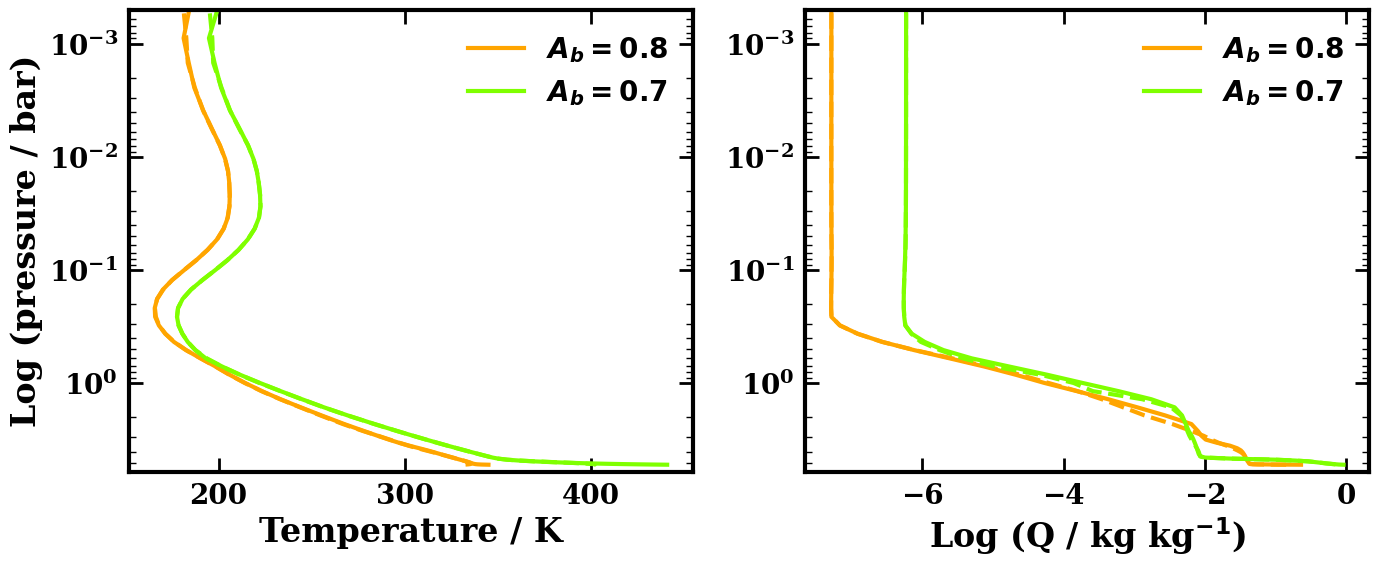}
    \caption{Pressure-temperature and pressure-moisture profiles at the substellar (solid lines) and antistellar (dotted) points for the 5 bar cases with imposed TOA albedos. There is increasing stratospheric temperature and moisture with increased instellation. The extent of the convective inhibition zone of the $A_b=0.7$ case is clear, and leads to temperatures well in excess of the boiling point of water at the surface, with a corresponding water vapour mixing ratio near unity.}
    \label{fig:r2_ab_5b_PTQs}
\end{figure*}

The pressure-temperature and pressure-moisture profiles for the 5 bar runs are shown in Figure \ref{fig:r2_ab_5b_PTQs}. The tropospheric temperatures are significantly cooler than in the 1 bar cases, a reflection of the lower OLR required to balance the incoming stellar radiation. As in the 1 bar cases, we are in a 'WTG' regime and so the tropospheric temperature differences are minimal. The only temperature gradients that can exist are driven by near-surface heating, and as there is less stellar energy reaching the near-surface, the surface energy gradients are weaker. The shape of the PT profiles are very similar to the 1 bar cases, but with a cooler tropopause at slightly higher pressures, the lapse rate following the moist adiabat below, and subsequently the increased surface pressure leading to higher surface temperature.
The moisture profiles follow the same qualitative pattern as the 1 bar cases, with the exception that there is a much greater region with shallow $q$ gradients in the $A_b=0.7$ case, reflecting widespread subsaturation near the surface.

\begin{figure*}
    \centering
    \includegraphics[width=0.95\textwidth]{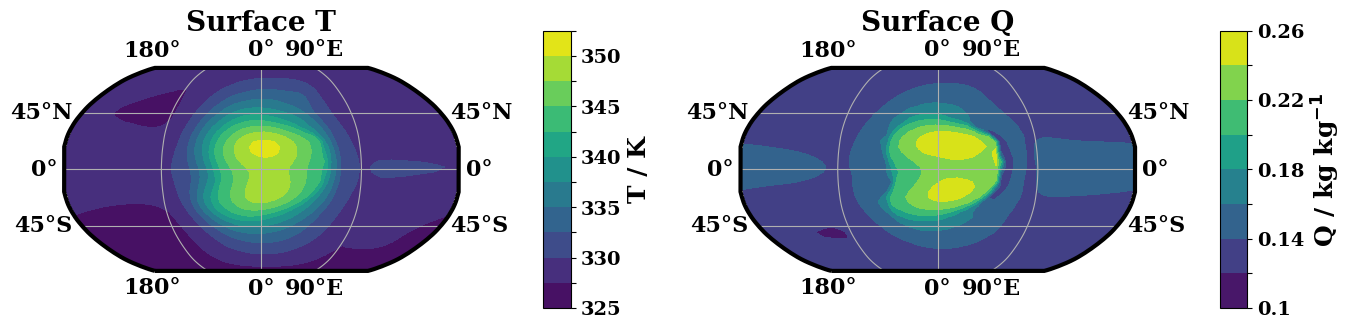}
    \caption{The surface temperatures and moisture mass fraction for the 5 bar case with imposed TOA albedo of 0.8. For a given surface temperature, the moisture mass fraction are much lower due to the increase in dry background pressure. The surface temperature differences are lower than the 1 bar cases, a reflection of less stellar radiation reaching the surface.}
    \label{fig:r2_ab_5b274w_TQSs}
\end{figure*}

We take a look at the surface temperatures and water fractions for the $A_b=0.8$ case specifically in Figure \ref{fig:r2_ab_5b274w_TQSs}. There are relatively small surface temperature differences of order 20-30K. This is in opposition to the $A_b=0.7$ case which is actively undergoing runaway greenhouse, where the convective inhibition layer increases the surface temperature gradient by almost 100K. The surface water fraction is still relatively low because of the significant dry pressure of the 5 bar atmosphere, although $q$ is higher than that of the 1 bar cases. As expected, because of the much greater greenhouse effect of the 5 bar atmosphere, the runaway greenhouse threshold happens for a lower instellation in the 5 bar cases than the 1 bar case.

Convective inhibition is still present in a largely familiar form, with a near saturated inhibition layer ($RH \approx 0.8$) in the substellar region. This is intermediate between the saturated and subsaturated 1 bar cases, and leads to a small amount of surface cloud cover, although most clouds happen higher up where some moist convection develops. This convection is relatively weak as diagnosed by its CAPE values, of order about 100 J kg$^{-1}$. This is likely because the reduced instellation and thicker atmosphere combine to produce a much reduced stellar forcing in the convective zone.

\begin{figure}
    \centering
    \includegraphics[width=0.48\textwidth]{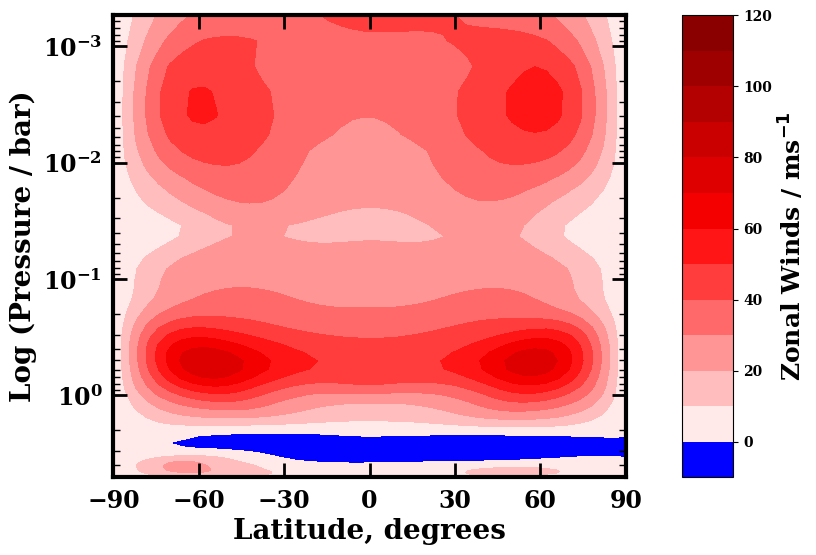}
    \caption{The zonally-averaged zonal winds for the 5 bar cases with imposed TOA albedo of 0.8. Some key differences are clear from the 1 bar cases: there are high-latitude cyclostrophic jets, but the lower section of the atmosphere below about 1.5-2 bar shows very little motion, in fact showing slow westward velocities.}
    \label{fig:r2_ab_5b274w_uavg}
\end{figure}

The dynamical structure does show differences from the 1 bar case. The most notable one is the presence of a band of retrograde motion in the lower atmosphere, as shown in the zonal wind structure of the $A_b=0.8$ case in Figure \ref{fig:r2_ab_5b274w_uavg}. This retrograde motion is not novel, and can be seen in e.g. \citet{Charnay2021}, \citet{Innes2022}, and \citet{Barrier2025}, although we note that the wind velocities below $\sim 2$ bar are small. We remain in a slow-rotator regime with equatorial Rossby waves. In this case, the tropospheric circulation is made up of mid-latitude zonal jets, similar to some of our 1 bar cases and to the 10 bar atmospheres in \citet{Innes2022} and \citet{Barrier2025}. The divergent circulation carries out the bulk of the dayside-nightside energy transport, although the jet and eddy components of the flow transport more heat than in the 1 bar cases - although usually they transport it in the opposite direction and so largely cancel each other out, similar to Figure \ref{fig:casestudy_uzonal_divsu}.

\section{Results III - Rayleigh scattering hazes}
\label{sec:r3_rs}
 
In reality, there needs to be a source of stratospheric reflection to generate a significant albedo, and this is most easily provided by high-altitude hazes. Simply modifying the incoming TOA stellar radiation, as done in the previous section, is a convenient but approximative method of imposing an albedo. The most common ways that incoming shortwave radiation is reflected are either from the surface or from clouds and hazes in the atmosphere. The atmospheres present here absorb enough radiation that only a small proportion of the incoming flux reaches the surface. For this same reason, clouds, especially low-lying ones, only have a small impact on the albedo as noted in \citet{Charnay2021}, \citet{Leconte2024}, and \citet{Jordan2025}. This leaves hazes as the most plausible source for an albedo. 

We use the term hazes to refer to radiatively active small particles suspended in the atmosphere \citep{Gao2021}. In opposition to clouds, they are not formed from condensed material; in the case of a cool exoplanet such as K2-18~b we would typically imagine them as photochemically produced \citep{Gao2021, Yu2020,Huang2024}. 
A full investigation into the effect of hazes would require an aerosol model establishing their formation, coagulation, vertical settling and transport, as well the optical properties of specific haze compositions, which are still in the process of being established \citep[e.g.][]{He2024}. Instead, here we aim to demonstrate the general effect that hazes have on the climate by parametrising the hazes as increased amounts of H$_2$ Rayleigh scattering, as in \citet{Piette2020}. This mimics a haze with only scattering and no absorption (like the extra-reflective soot hazes in e.g. \citet{Malsky2025}), and with the Rayleigh scattering wavelength dependence of $\sigma \propto \lambda^{-4}$.

To arrive at our the total haze optical depths, we increase the total Rayleigh scattering  cross-section by a factor $E$, such that the haze optical depth of a layer $d\tau_h(\lambda) = \sigma_{ray,H_2}(\lambda) n_{H_2} E$. As described in Section \ref{sec:methods}, we assume a uniform horizontal haze distribution, but we reduce our enhancement factor $E$ at pressures below 100 mbar. We note that this haze distribution is an approximate method of representing horizontal and vertical variations in haze abundances, and more complex models may lead to different results.

We run simulations with enhancement factors of 100$\times$, 1000$\times$, and 10,000$\times$ for 1 bar and 5 bar atmospheres. We also include a 500$\times$ enhancement simulation for a 1 bar atmosphere and a 4000$\times$ enhancement simulation for a 10 bar atmosphere. This is to further constrain the runaway greenhouse threshold albedo in the 1 bar case, and to find an upper bound on the threshold albedo in the 10 bar case. The value for the 10 bar case was chosen based on extrapolating the threshold albedos for the 1 and 5 bar cases, choosing a value that was very likely to lead to stable climate whilst also being reasonably close to the runaway threshold. Indeed, the resulting temperatures we find are close to a being in runaway. The instellation is kept at 1368 Wm$^{-2}$ as appropriate for K2-18~b, and all other model parameters are kept the same as the cases in Section \ref{sec:r2_ab}, i.e. matching the values in Table \ref{tbl:k218b_params}. We calculate the resulting albedo by seeing what fraction of incoming stellar radiation is outgoing at the top of the atmosphere. In the following section we discuss the constraints on the amounts of hazes necessary to achieve a stable climate state with a liquid water ocean, and also describe the effect of hazes on the temperature and climate structure more generally.

\begin{table*}
\centering
\begin{tabular}{cccccc}%{|c|c|c|c|c|c}

\hline \hline
\textbf{Run Name} & \textbf{Surface Pressure / bar}  & \textbf{Rayleigh Scattering factor} & \textbf{Induced albedo} & \textbf{Run time / days} & \textbf{Converged?} \\
\hline
1b-100x & 1 & 100 & 0.17 & 18,000 & \color{red} $\times$ \\
1b-500x & 1 & 500 & 0.27 & 35,000 & \color{green} \checkmark \\
1b-1000x & 1 & 1000 & 0.34 & 35,000 & \color{green} \checkmark \\
1b-10000x & 1 & 10000 & 0.55 & 35,000 & \color{green} \checkmark \\
\hline
5b-100x & 5 & 100 & 0.22 & 16,000 & \color{red} $\times$ \\
5b-1000x & 5 & 1000 & 0.35 & 55,000 & \color{green} \checkmark \\
5b-10000x & 5 & 10000 & 0.56 & 55,000 & \color{green} \checkmark \\
\hline
10b-4000x & 10 & 4000 & 0.48 & 75,000 & \color{green} \checkmark \\
\hline \hline
\end{tabular}
\caption[Simulation details of Rayleigh scattering suite]{Simulation details of the second suite of runs with enhanced Rayleigh Scattering. A run which has converged is one which has reached a stable climate state, the ones which have not are in a runaway greenhouse state.}\label{tbl:rs_runs}
\end{table*}

\subsection{Threshold Albedo for Runaway Greenhouse}

The albedos we find for our various cases are shown in Table \ref{tbl:rs_runs}. We confirm that for high enough amounts of Rayleigh scattering, a stable climate state can be found. The effective albedo at which this happens is notably lower than the threshold albedo seen in Section 3. For the 1 bar atmospheres, the critical Rayleigh scattering enhancement lies between 400x and 500x, corresponding to a threshold albedo between 0.25 and 0.27. For the 5 bar atmospheres, the critical Rayleigh scattering enhancement is between 100 $\times$ and 1000 $\times$ and is likely close to 1000 $\times$, corresponding to a threshold albedo of about 0.35. For the 10 bar atmosphere, an albedo of 0.48 allows a stable climate, with the threshold albedo likely slightly smaller than this.

The difference between the TOA and the haze threshold albedo is because hazes have a greater impact on temperature structure than just increasing the albedo. By increasing scattering throughout the atmosphere, and especially above 100 mbar following our vertical parametrisation, they increase shortwave absorption higher up in the stratosphere, and correspondingly reduce it lower down in the atmosphere as there is less stellar flux reaching those regions \citep{Piette2020, Leconte2024}. This leads to a hotter stratosphere and a correspondingly cooler troposphere, resulting in lower surface temperatures than if the same albedo was imposed as a TOA flux reduction like in Sections \ref{sec:casestudy} and \ref{sec:r2_ab}.

\begin{figure*}
    \centering
    \includegraphics[width=0.95\textwidth]{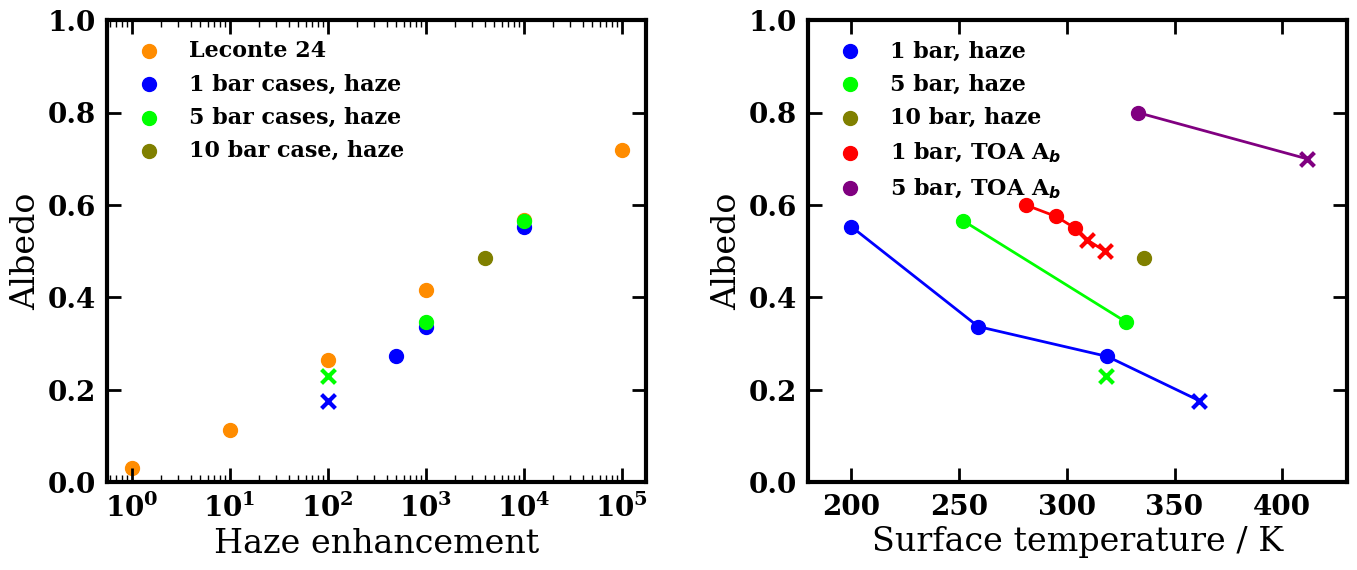}
    \caption{The summary of the haze cases. Crosses denote cases which do not have stable Hycean climates. Left: the resulting albedo for a given Rayleigh scattering enhancement. We find similar but slightly lower albedo values than \citet{Leconte2024}. This difference is likely caused by our modified haze vertical profile, with reduced scattering and so reduced reflection from the lower atmosphere. Right: albedo vs average surface temperature of our runs, including the imposed TOA albedo cases, showing the increased ability of hazes to reduce the surface temperature. We note that the surface temperature of the runaway greenhouse cases represent out-of-equilibrium conditions, and that the 5 bar, 100 $\times$ case was terminated early, explaining its moderate temperatures.}
    \label{fig:r3rs_constraints}
\end{figure*}

We show the overall runaway greenhouse constraints for our Rayleigh scattering cases in Figure \ref{fig:r3rs_constraints}, alongside the haze-albedo contraints from \citet{Leconte2024}. For the same amount of Rayleigh scattering, we find a somewhat lower albedo: this is a result of the non-uniform vertical haze profile. We see that stable Hycean climates are possible for an albedo of 0.27 in the 1 bar case. This is drastically lower than found in previous work \citep{Innes2023,Leconte2024} and shows the efficiency of hazes at reducing surface temperatures.

There is debate over what levels of hazes are possible given observations. There is some evidence for hazes in the K2-18b JWST observations, though their precise properties are relatively unconstrained. However, strong hazes can mute the clear methane and carbon dioxide features seen. We discuss the interpretation of the haze constraints in Section \ref{sec:summary}.

\subsection{Differences in climate state and dynamics}

\begin{figure*}
    \centering
    \includegraphics[width=0.95\textwidth]{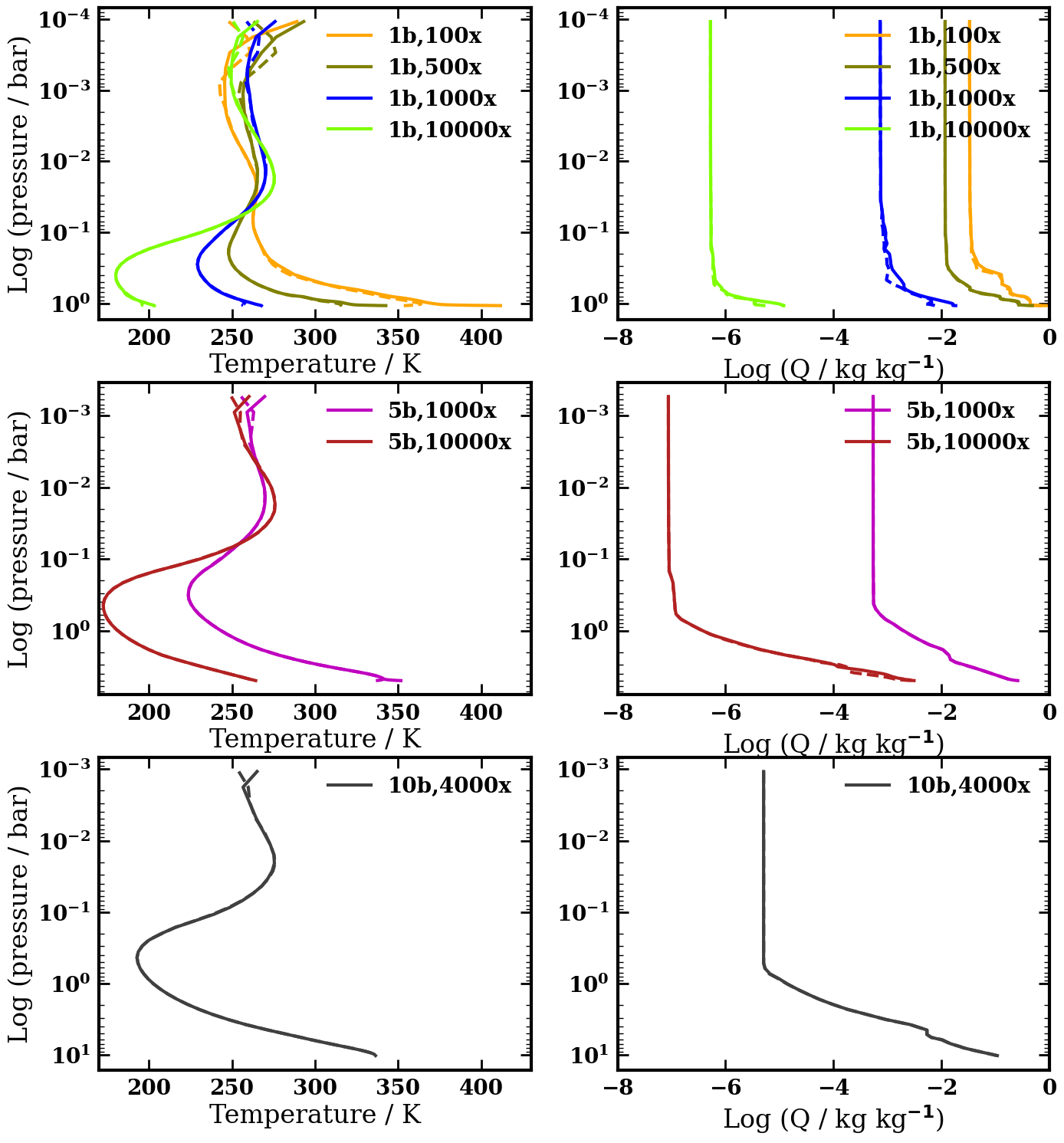}
    \caption{Pressure-temperature and pressure-moisture profiles at the antistellar and substellar points for the 1, 5, and 10 bar cases with increased Rayleigh scattering. Solid (dotted) lines denote substellar (antistellar) point profiles. We note that the 1 bar, 100 $\times$ haze case is in a runaway greenhouse state. We note that we do not include the 5 bar, 100$\times$ case. This was deemed to be in a runaway greenhouse state based on the behaviour of the 1 bar, 100 $\times$ case, and was terminated early to save computational resources. The influence of hazes on the temperature profiles is apparent: the stratospheric temperatures are close to the equilibrium temperature of the planet (250-300K, depending on the albedo), but there is a large temperature inversion  before a tropopause at 0.2-0.5 bar. At greater pressures than that, the lapse becomes positive again, although in for increased hazes and / or thinner atmospheres the surface temperature is often lower than the peak stratospheric temperature. The moisture profiles follow expected values, with extreme cold trapping in the strong haze cases.}
    \label{fig:r3rs_PTQs}
\end{figure*}

The overall climate state and dynamics are very similar to that of the imposed Bond albedo cases. Despite this, there are some differences. The first and most important is that of the vertical profiles of temperature and subsequently of moisture,  as shown in Figure \ref{fig:r3rs_PTQs}. In all our cases, we end up with stratospheric temperatures of at least 250K, necessary for the OLR to match the incoming stellar radiation. Below that is a significant temperature inversion. This is larger than that seen in the imposed TOA bond albedo cases: almost 100K in the case of the 10,000$\times$ scattering cases. This has corresponding impacts on the moisture profiles: as the tropopause is at a correspondingly greater temperature, there is more stratospheric moisture for a given surface temperature. There is still moist convection in the substellar region, although the lower stellar flux reaching the troposphere means that more of it is in radiative equilibrium and the convection zone is shallower. 

The dynamical structure of these runs largely stay in the slow rotator configuration, although some have mid-latitude jets. There are two main factors which we expect to lead to differences in dynamical structure in these cases. Compared to the simulations presented in Section \ref{sec:r2_ab}, there is much more energy absorbed in the stratosphere and somewhat less in the lower troposphere, and the lapse rates throughout are much lower. This leads to a significant increase in the static stability of the atmosphere, raising the Rossby number and disfavouring a fast rotator state. The optical depths of the atmospheres are also in general greater, which pushes the dynamically active region of the atmosphere to slightly lower pressures and reduces the horizontal temperature differences. The mid-latitude jets thus do not seem to be due increased baroclinicity, but they may thus be due to the greater instellation, or to the longer radiative timescale of the atmosphere - while the simulations have a stable thermal structure, it is possible that a longer runtime would lead to changes in the dynamical structure, perhaps similarly to \citet{Wang2020}.

\section{Summary and Discussion}
\label{sec:summary}

%\subsection{Summary}

We report the first GCM simulations of Hycean planets. We investigate their climate and dynamical structure, focusing on the range of dynamical states present, the presence and behaviour of convective inhibition, and the location of the runaway greenhouse threshold.
We conduct GCM simulations of K2-18b assuming a surface liquid water ocean at pressures of 1 and 5 bar. We vary the energy input first by varying the imposed top of atmosphere albedo, and then by changing the H$_2$ Rayleigh scattering enhancement factor, varying it from values of 100x to 10,000x.

We draw the following key conclusions:
\begin{itemize}
    \item When varying the imposed TOA albedo, the runaway greenhouse threshold lies at an albedo of 0.525 and 0.55 for a 1 bar surface pressure, and between 0.7 and 0.8 for a 5 bar surface. This is consistent with the work of \citet{Leconte2024}.
    \item When varying the Rayleigh scattering enhancement factor, the threshold enhancement for the 1 bar cases lies between 400x and 500x, corresponding to an albedo of 0.25 to 0.27. For the 5 bar cases, the threshold albedo is about 0.35. We also run a 10 bar case, finding that an albedo of 0.48 allows for stable conditions. These albedo constraints are lower because scattering hazes are more efficient at reducing surface temperatures than a TOA flux reduction.
    \item These enhancement factors are not ruled out by observations. From the climate point of view, a Hycean scenario for K2-18b remains eminently plausible and chemical modelling must be used to differentiate between possible scenarios. %These enhancement factors are consistent with observations, and in fact there must be at least some albedo to explain the non-detection of H$_2$O.
    \item A Hycean K2-18b is broadly in a slow-rotating dynamical state, characterised by low free atmosphere temperatures differences, divergent circulation dominating heat transport, and either an equatorial super-rotating jet or mid-latitude cyclostrophic jets.
    \item In the substellar zone, we find near-surface convective inhibition that broadly matches the predictions from \citet{Seeley2025} at moderate temperatures, but a subsaturated inhibition zone at greater temperatures.
\end{itemize}

\subsection{Dynamical structure of Hycean planets}

We present the first look at the dynamical structure of a Hycean atmosphere. This is largely in line with theoretical expectations and previous GCM work on similar planets. We see weak horizontal temperature gradients in the free atmosphere due to the `WTG' regime K2-18 b finds itself in. However, near-surface friction still allows for significant surface temperature variations, especially in thinner atmospheres, and convective inhibition amplifies these effects on the dayside. 

We mostly see a slow-rotator dynamical structure \citep[e.g.][]{Haqq-Misra2018}. There are equatorial Rossby waves, an eastwards shift of the maximum heating location, a general dayside-nightside overturning circulation, and strong upper-atmosphere super-rotation. The rotational component of the wind dominates the wind field, although the divergent circulation is responsible for the most of the dayside-nightside energy transport. 

The jet structure is slightly more complex. In some cases  - shallower and cooler atmospheres - we see a single broad tropospheric equatorial jet, as is standard in slow rotator cases. In other cases, however, we find mid-latitude zonal jets with axial angular momentum peaking in the mid-latitudes. In a few of these cases, particularly with higher instellation, there are extra-tropical Rossby waves instead of equatorial ones, placing these cases in a `fast rotator' regime instead. These differences seem to be due to greater substellar point-pole temperature differences - in some cases driven by enhanced convective inhibition - which enhance the formation of cyclostrophic jets. We note that conventional scaling relations such as the dimensionless lengths in Equations \ref{eq:l_ro} and \ref{eq:l_rh} have somewhat limited predictive power, particularly when tropospheric averages are used.

\subsection{Comparison with previous work}

An important question in this work was to see how the runaway greenhouse predictions from our 3D simulations compared to that from 1D models, specifically those of \citet{Innes2023} and \citet{Leconte2024} (the latter informed by high-resolution 3D models). We find that our results are consistent with those of \citet{Leconte2024} when we use the same parametrisation to induce an albedo. They do differ from those of \citet{Innes2023}, largely because their 1D model did not include small-scale turbulence as a heat transport mechanism. This indicates that 1D modelling, with the appropriate physical mechanisms parametrised, is a reasonably accurate way to model the atmosphere of such planets. This is because of the slow rotation of the planet which places it in a WTG regime \citep{Pierrehumbert2016}, and the fact that the photosphere is mostly located in the free troposphere and stratosphere. The tropopause temperatures in those regions are globally uniform and set by the planet-averaged OLR required. The substellar pressure-temperature profile then follows the moist adiabat to the surface, meaning that the overall P-T profile here is the same as would be found in a 1D model with a planet-averaged irradiation. This would not necessarily be the same in faster rotating atmosphere with larger horizontal temperature differences, either day-nightside or pole-equator. The weak influence of clouds on the albedo also contributes to making 1D models more accurate, in comparison to terrestrial tidally-locked planets where thick dayside clouds can substantially affect the IHZ \citep{Yang2014,Kopparapu2016}. However, in our Section \ref{sec:r3_rs} we assumed horizontally uniform hazes, and it is not clear if this is a valid assumption: it is possible that dayside/nightside/terminator differences in haze profile would render 1D models unsuitable for studying these atmospheres.

As we are parametrising moist convection and moist convective inhibition, which are still relatively poorly understood phenomena, we are also interested in seeing how convective inhibition develops in a global setting. We find that some of the qualitative and quantitative behaviour, including the near-surface lapse rates, matches the high-resolution findings from \citet{Seeley2025}. However, we find a variety of other convective states: a subsaturated convective surface layer with no convective inhibition in our colder cases, and a subsaturated inhibition zone in hotter cases. Taken together, all this shows the utility of a hierarchy of models (high-resolution CPM, GCM, and 1D models) in exploring climate features as efficiently and accurately as possible, and of the necessity for further investigations into these effects, particularly in hotter conditions near the runaway greenhouse limit.

\subsection{Albedo requirements}

A key finding from this work are constraints on the albedos required for a stable Hycean K2-18b climate. We first use a rough parametrisation for the albedo of reducing the incoming flux arriving at the top of the atmosphere. We find that the boundary lies between an albedo of 0.525 and 0.55 for a 1 bar surface pressure and between 0.7 and 0.8 for a 5 bar surface pressure, consistent with \citet{Leconte2024}.  We note that this TOA albedo method is an imperfect approximation of the climate effects of real albedo sources. It serves to reduce the surface temperature - and delay the initialisation of a runaway greenhouse state - by reducing the photospheric temperature, but otherwise leaves the atmospheric lapse rates very similar. 

In reality, clouds, hazes, or a reflective surface are needed to actually generate the required albedo \citep{Piette2020}. Some options are unfeasible for a Hycean K2-18b. Relatively little radiation reaches the surface, and so an increased surface albedo makes little difference to the total albedo. Tropospheric clouds have a similarly limited impact as they are too low down in the atmosphere to make a significant difference \citep{Jordan2025}. That leaves hazes as the most likely source of an albedo.

We use a simple haze parametrisation of increasing the H$_2$ Rayleigh scattering by a given factor, which we vary between either 100$\times$, 1000$\times$, or 10000$\times$ (with 1x representing a clear atmosphere). This does not investigate all possible haze behaviour: while we can change the enhancement factor, we do not change the wavelength dependence of the Rayleigh scattering, so $\sigma \propto \lambda^{-4}$. We also treat the hazes as horizontally but not vertical uniform, and reduce the haze enhancement at high pressures. We find that hazes are effective at reducing surface temperatures well beyond what their induced albedo would suggest. This can be understood by considering that their opacity is much greater in the incoming shortwave than the outgoing longwave radiation. This reduces the radiative lapse rate, or even leads to temperature inversions \citep{Madhusudhan2019}. In our case, increasing hazes - on top of their role in increasing the albedo - leads to a warmer stratosphere and a correspondingly cooler troposphere, reducing surface temperatures and preventing a runaway greenhouse effect.

We find that for the 1 bar atmospheres, cases with Rayleigh scattering enhancement factors of 400$\times$ enter a runaway greenhouse state while the 500$\times$ cases maintain a stable climate, constraining the runaway greenhouse threshold albedo to between 0.25 and 0.27. For the 5 bar cases, cases with Rayleigh scattering enhancement factors of 100$\times$ enter a runaway greenhouse state while the 1000$\times$ cases maintain a stable climate, with the 5 bar threshold correspondingly slightly below 0.35. For a 10 bar atmosphere, a 4000$\times$ case with a corresponding albedo of 0.48 is stable. While we have only run one 10 bar simulation, this case is not far from the runaway threshold as diagnosed by its average surface temperature. Therefore, the threshold albedo for a 10 bar atmosphere is likely to be slightly lower than 0.48. We note that these required albedos are substantially lower than those calculated in the reduced TOA flux cases. This is the first of two key points that lead to our conclusions differing from \citet{Leconte2024}: given on how the albedo is implemented, we find that the runaway greenhouse threshold albedo changes.

\subsection{Feasibility of this albedo}

 \begin{figure*}
    \centering
    \includegraphics[width=0.95\textwidth]{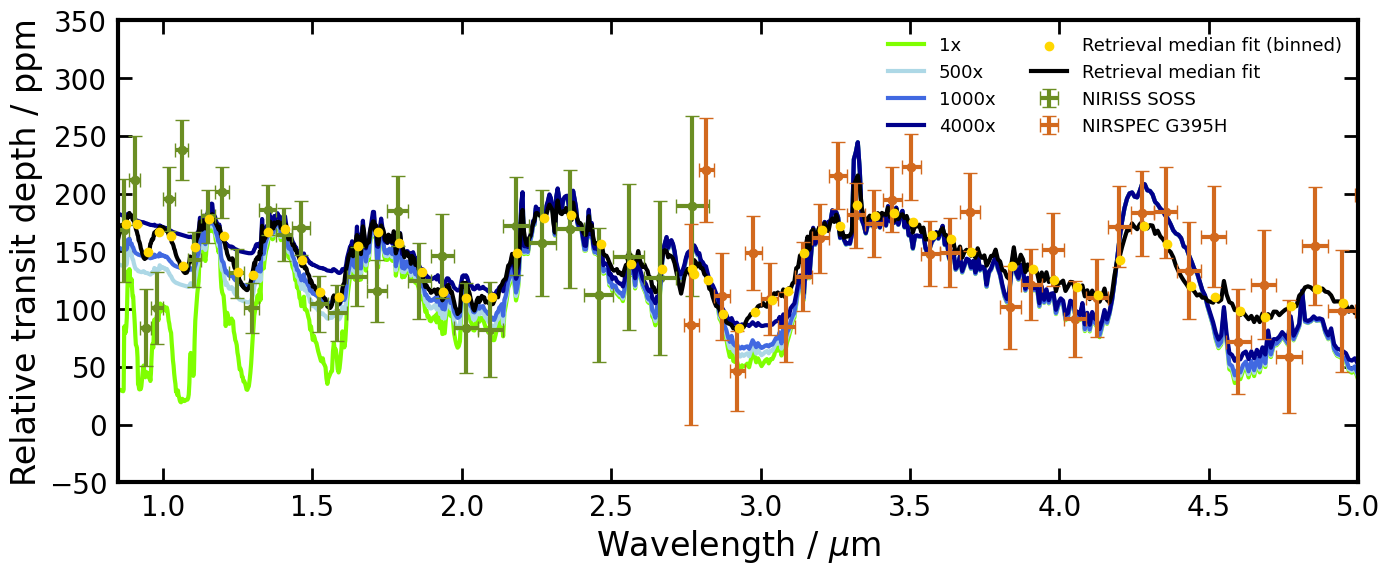}
    \caption{Simulated transit spectra alongside the JWST data. We show the NIRISS SOSS and NIRSPEC G395H spectra from \citet{Madhusudhan2023b}, alongside the median retrieved fit and the median model binned down to match the observations. We also show transit spectra, simulated using the PSG \citep{Villanueva2018}, of our GCM runs with various of hazes. As we did not run GCMs for haze enhancement values of 1$\times$, we the GCM outputs from the 1 bar, 1000$\times$ scattering case and reduce the haze abundances accordingly. We note that a retrieval is necessary to quantitatively assess the haze properties, but the haze properties considered here are qualitatively in agreement with the spectra, with the likely exception of the 1x and 4000x scattering cases}.
    \label{fig:s6_ts_obs}
\end{figure*}

The obvious next question is whether a level of haze generation needed to produce these amounts of hazes is possible. We think it is. Photochemically modelling of thin H$_2$ atmospheres overlying a liquid water ocean \citep{Madhusudhan2023a, Huang2024} indicate that the condensation of water and resulting high C/O ratio would lead to the formation of large amounts of longer chain hydrocarbons, which would be expected to go on to form hazes. A more complicated question is whether the hazes formed would be consistent with observations. \citet{Leconte2024} argued that the amount of hazes required for stability would flatten the spectral features to a degree inconsistent with observations. However, they appear to have taken the required haze enhancement factor as the enhancement factor required to generate the necessary albedo as calculated by the TOA flux reduction method, whereas we have seen that the two methods have different impact on the surface temperatures and, hence, the runaway greenhouse threshold. \citet{Leconte2024}'s Figure 11 can thus be interpreted as showing that an enhancement factor of a few hundred would still leave large enough features to be consistent with observations, which is almost enough of an enhancement factor to lead to stable surface conditions in the 1 bar case. 

The second key difference with \citet{Leconte2024} which affects our conclusions is a different approach to evaluating the compatibility of a given haze enhancement with the observations. In Figure \ref{fig:s6_ts_obs}, we show simulated transit spectra for a variety of haze enhancements, calculated by the Planetary Spectrum Generator \citep{Villanueva2018} using our GCM outputs. Overplotted on these are the NIRISS SOSS and NIRSPEC G395H transit spectra from \citet{Madhusudhan2023b}, using the JExoRes pipeline \citep{Holmberg2023}. We also add the median retrieved spectrum from \citet{Madhusudhan2023b}, along with the median spectrum binned to match the transit spectra. A few features stand out. First of all, the impact of the increased hazes is apparent from about 1.5 $\mu$m bluewards, leading to increased transit depths and flattened spectra. Compared to a haze free atmosphere, the depth of the $\approx 1.4 \mu$m methane feature is approximately halved for a 1000$\times$ haze enhancement. Crucially, the different haze enhancements shown are all consistent with the data within the $1\sigma$ uncertainty on average. There is enough scatter in the data and the uncertainties are large enough that none of the spectra are obviously preferred to any others, and allow for at least moderate hazes. The one partial exception to this is the $\geq$4000$\times$ haze enhancement required to stabilise a 10 bar atmosphere, the spectrum for which is inconsistent with some of the data points below 1.6 $\mu$m to within their 1-$\sigma$ uncertainties. Super-Rayleigh scattering slopes and/or dayside-terminator haze differences would then need to be invoked to allow a 10 bar Hycean atmosphere. 

We also note that comparing the height of spectral features is only a qualitative assessment of the amount of hazes, and that atmospheric retrievals are needed for statistically robust inferences. The median retrieval (shown in yellow in the figure), does in fact find some evidence for hazes \citep{Madhusudhan2023b}. The best fit values correspond to an H$_2$ Rayleigh enhancement factor of $\approx10^{8.6}$ at $0.35 \mu m$, with a power law slope of $\approx -11$. These do have large uncertainties, and the Rayleigh enhancement factors we use are consistent with these retrieved quantities. We note that the best fit values do in fact give significant amounts of hazes as noted in \citet{Jordan2025}, who found they led to an albedo in the range of 0.42-0.49, similar to our threshold albedo for the 10 bar case. Additionally the scattering slope of $\approx -11$, achievable for example by a vertical gradient in haze opacity \citep{Ohno2020}, would be expected to lead to a stronger version of the decreased lapse rate and/or temperature inversion argument we have outlined above, further reducing surface temperatures.

Overall, while the transmission spectrum does not allow us to claim the existence of hazes, it should not be used to argue against some amount of haze presence either. Future optical and NIR observations of K2-18b may be able to tell us more about the presence and type of any hazes, by constraining the size of the methane features and the scattering slope at wavelengths shorter than 1 $\mu m$.

The lack of detected water in K2-18b's transmission spectrum is likely to impose surprisingly stringent constraints on the temperature profile. The 2-$\sigma$ upper-detection limit is for a volume mixing ratio of 10$^{-3.06}$ \citep{Madhusudhan2023b}. Regardless of K2-18b's internal structure and atmospheric composition, it is likely that saturation and water condensation happens at some point in the atmosphere. The only other alternative is for a deep atmosphere with a very low H$_2$O volume fraction, although the C/O ratio must be pushed to extremely high values (e.g. C/O = 3.5 in the case of \citet{Schmidt2025}) for this to happen. We note that this is only possible if no CO$_2$ is detected - in the deep atmosphere the presence of H$_2$O is favoured over CO$_2$ \citep{Cooke2024}, so the presence of CO$_2$ would imply substantial H$_2$O at depth which must condense out.

This requirement for condensation means that the temperature profile must - at some point - be cooler than the dewpoint temperature of an atmosphere with a water VMR of  10$^{-3.06}$. This is likely to happen at the tropopause, and atmospheric temperatures can increase above that, but we will still get a dry stratosphere as seen on Earth and in our cases, for example in Figure \ref{fig:r3rs_PTQs}. As expected and as others have found \citep{Cooke2024, Huang2024}, a significant albedo is required for the H$_2$O VMR fraction to match observations. In our case, the 1 bar 100x and 500x cases are disfavoured, and the 1 bar and 5 bar 1000x cases are only just in agreement with the retrieved constraints. In fact, we end up in a position where - given our assumptions on the form of the albedo-generating haze - all shallow Hycean atmospheres which are cool enough to satisfy the non-detection of water vapour are also cool enough to maintain a liquid water ocean. Of course, the same is not true of thicker atmospheres, where the required haze amounts would increase substantially.

These arguments all mean that the odds of a Hycean K2-18b having a stable atmosphere are much higher than previously thought. There must be some albedo, and given the albedo constraints there is a good chance that this albedo is high enough for a Hycean climate to be stable. This is not too say that K2-18b has to be Hycean - the internal structure is degenerate and other options may match the chemical signatures, for example a super-critical water layer. But it is a very real possibility.

\section*{Acknowledgements}

 E.B. and N.M. acknowledge support from the UK Research and Innovation (UKRI) Frontier Grant (EP/X025179/1, PI: N. Madhusudhan) towards the doctoral studies of E.B. This work was performed using resources provided by the Cambridge Service for Data Driven Discovery (CSD3) operated by the University of Cambridge Research Computing Service (www.csd3.cam.ac.uk), provided by Dell EMC and Intel. 

 %We acknowledge - Greg for discussion re hazes, Eric Wolf for discussions re ExoCAM, Savvas for discussions re hazes

%%%%%%%%%%%%%%%%%%%%%%%%%%%%%%%%%%%%%%%%%%%%%%%%%%
\section*{Data Availability}

 The default ExoCAM GCM which we build off is available at https://github.com/storyofthewolf/ExoCAM. Models outputs are available on reasonable request from the paper authors.



\bibliographystyle{mnras}
\bibliography{paper} % if your bibtex file is called example.bib

%\begin{thebibliography}{99}
%\end{thebibliography}

%%%%%%%%%%%%%%%%%%%%%%%%%%%%%%%%%%%%%%%%%%%%%%%%%%

%%%%%%%%%%%%%%%%% APPENDICES %%%%%%%%%%%%%%%%%%%%%

% \appendix

% \section{Some extra material}

% If you want to present additional material which would interrupt the flow of the main paper,
% it can be placed in an Appendix which appears after the list of references.

%%%%%%%%%%%%%%%%%%%%%%%%%%%%%%%%%%%%%%%%%%%%%%%%%%

% Don't change these lines
\bsp	% typesetting comment
\label{lastpage}
\end{document}